\documentclass[a4paper,11pt]{article}
\pdfoutput=1

\usepackage{amsmath}
\usepackage{graphicx}
\usepackage{amsfonts}
\usepackage{amssymb}
\usepackage{tikz}
\usepackage{color}
\usepackage{comment}

\newcommand{\bc}{\begin{center}}
\newcommand{\ec}{\end{center}}
\def\ba#1{\begin{array}{#1}\displaystyle}
\newcommand{\ea}{\end{array}}

\newcommand{\beq}{\begin{equation}}
\newcommand{\eeq}{\end{equation}}
\newcommand{\beqs}{\begin{equation*}}
\newcommand{\eeqs}{\end{equation*}}
\newcommand{\beqa}{\begin{eqnarray}}
\newcommand{\eeqa}{\end{eqnarray}}
\newcommand{\beqas}{\begin{eqnarray*}}
\newcommand{\eeqas}{\end{eqnarray*}}

\newcommand{\n}{\nonumber\\}
\newcommand{\bi}{\begin{itemize}}
\newcommand{\ei}{\end{itemize}}

\def\lt#1{\left#1}
\def\rt#1{\right#1}
\def\t#1{\tilde{#1}}

\def\b#1{\bar{#1}}
\def\frc#1#2{\frac{#1}{#2}}
\newcommand{\p}{\partial}

\newcommand{\bra}{\langle}
\newcommand{\ket}{\rangle}
\newcommand{\Z}{{\mathbb{Z}}}

\newcommand{\R}{{\mathbb{R}}}
\newcommand{\C}{{\mathbb{C}}}

\newcommand{\dd}{\mathrm{d}}
\newcommand{\ii}{{\mathrm{i}}}

\newcommand{\Or}{{\cal O}}

\newcommand{\ep}{\epsilon}
\newcommand{\varep}{\varepsilon}

\newcommand{\Tr}{{\rm Tr}}

\newcommand{\sta}{{\rm sta}}

\begin{document}

\begin{titlepage}

\begin{center}
{\Large {\bf Conformal field theory out of equilibrium: a review}

\vspace{1cm}

Denis Bernard${}^{\clubsuit}$ and Benjamin Doyon${}^{\spadesuit}$

}

\vspace{0.5cm}
{\small ${}^{\clubsuit}$ Laboratoire de Physique Th\'eorique de l'Ecole Normale Sup\'erieure de Paris, \\
CNRS, ENS \& PSL Research University, UMPC \& Sorbonne Universit\'es, France.}\\
{\small ${}^{\spadesuit}$ Department of Mathematics, King's College London, London, United Kingdom.}\\

\end{center}

\vspace{1cm}

\noindent We provide a pedagogical review of the main ideas and results in non-equilibrium conformal field theory and connected subjects. These concern the understanding of quantum transport and its statistics at and near critical points. Starting with phenomenological considerations, we explain the general framework, illustrated by the  example of the Heisenberg quantum chain. We then introduce the main concepts underlying conformal field theory (CFT), the emergence of critical ballistic transport, and the CFT scattering construction of non-equilibrium steady states. Using this we review the theory for energy transport in homogeneous one-dimensional critical systems, including the complete description of its large deviations and the resulting (extended) fluctuation relations. We generalize some of these ideas to one-dimensional critical charge transport and to the presence of defects, as well as beyond one-dimensional criticality. We describe non-equilibrium transport in free-particle models, where connections are made with generalized Gibbs ensembles, and in higher-dimensional and non-integrable quantum field theories, where the use of the powerful hydrodynamic ideas for non-equilibrium steady states is explained. We finish with a list of open questions. The review does not assume any advanced prior knowledge of conformal field theory, large-deviation theory or hydrodynamics.

\vfill

{\ }\hfill 
\today

\end{titlepage}

{\setlength{\parskip}{0pt plus 1pt}
\tableofcontents}

\section{Introduction}

Quantum physics out of equilibrium has received a lot attention recently. Experimental and theoretical studies have given rise to a landscape of results, and it is important to attempt to extract the physically relevant and general concepts. 

One possible route is to investigate dynamical properties: responses to excitations or pulses, approaches to steadiness and thermalization. The questions of the dynamics under a  time-dependent local hamiltonian, often where a coupling is suddenly switched on or off (so-called ``quantum quenches'') \cite{quench}, and the related question of equilibration and thermalization, have led to a very large amount of work recently. A central idea in the context of thermalization is that of the eigenstate thermalization hypothesis \cite{ETH}, see also the works \cite{Tas98,rigol1,RGE12,SKS13, MAMW15,Dtherm} and the reviews \cite{thermali}. Recent progress has been made especially concerning the effect of integrability, where generalized Gibbs ensembles emerge \cite{GGE} (the concept has beed refined in \cite{Dtherm,GGEfinal}), as shown in many examples \cite{GGEfinal,GGEex}. Reviews are dedicated to aspects of this subject in the present volume \cite{CCreview,EFreview,Pr_thisreview}.

Another route is to study steady properties: currents carried by steady flows, responses to external drives. These physical phenomena have been investigated for a long time in the context of mesoscopic transport, see e.g. the books \cite{meso}. Because, in steady states, properties do not change with time, the study of such phenomena has opened, at least in classical systems \cite{classical}, the door to thermodynamic-like descriptions, to a deeper understanding of fluctuations, and to the establishment of general principles, see e.g. \cite{Derrida-revue,MFT,Jona,Bodineau-Derrida}. Reviews in the present volume discuss aspects of quantum steady states in integrable models \cite{Pr_thisreview,VMreview}.

In all cases, out-of-equilibrium phenomena display nonzero macroscopic or mesoscopic flows. Dynamical and steady properties are of course related to each other via the dynamics giving rise to these flows. An important properties of these flows is that they are not time-reversal invariant, and indeed out-of-equilibrium phenomena are often fundamentally characterized by states or ensembles of trajectories where time-reversal invariance has been broken. One of the aims of out-of-equilibrium physics is to decipher statistical properties of such flows and the underlying principles governing them, beyond the linear response approximation.

The goal of this manuscript is to review recent progress, based on the works \cite{BD2012,BD-long}, in constructing and analyzing non-equilibrium phenomena within conformal field theory and perturbation thereof. This may have direct applications to transport phenomena in critical and mesoscopic systems. It also may provide workable and useful examples of steady states far from equilibrium where important notions in non-equilibrium physics can be studied with precision, including fluctuation relations of the Cohen-Gallavotti \cite{fluctu} and Jarzynski type \cite{Jarz} and extensions thereof (see \cite{Espo} for a review), Onsager reciprocal relations and emerging hydrodynamics. This could guide us in deciphering properties and principles underlying out-of-equilibrium quantum physics in extended and interacting many-body systems. 

The review is organized as follows. In section \ref{sectbasics} we overview basic aspects of mesoscopic physics, in order to develop a useful context in which the results reviewed can be interpreted. In section \ref{sectgen}, we explain the general framework: we describe the partitioning approach for generating quantum steady states, consider a general linear-response theory analysis of this approach and describe an explicit quantum example, that of the Heisenberg chain. In section \ref{sectCFT}, we develop one-dimensional non-equilibrium conformal field theory from basic principles, explaining the notion of universal steady-state limit and how chiral factorization leads to the exact non-equilibrium state and its energy current. In section \ref{sectfluctu}, we continue the study of the energy current by concentrating on the scaled cumulant generating function, developing the large-deviation theory. In sections \ref{sectcharge} and \ref{sectdefects} we extend some of these results to charge transport and to the presence of defects. Finally, in section \ref{sectbeyond}, we show how to extend the ideas and some of the results beyond one-dimensional conformal field theory, making connections with other non-equilibrium ensembles studied in the linterature and developing the notion of hydrodynamics.

\section{Mesoscopic electronic transport: basics} \label{sectbasics}

This review is concerned with examples of systems where ballistic, coherent quantum transport exists. This is a situation which is typical of (electronic) transport in mesoscopic systems, but also, as we will explain, of critical many-body quantum systems because of emerging behaviours. Coherence also impacts thermal transport in mesoscopic system leading to a quantum of thermal conductance, theoretically predicted \cite{conductance} and experimentally measured \cite{Schwab,Pierre_et_al}.   In order to put the ideas and results into context, we start with a general phenomenological description of coherent transport (see e.g. \cite{meso} and references therein), and a brief review of some fundamental aspects of the linear-response theory associated to ballistic transport.

\subsection{Elementary phenomenology}\label{ssectpheno}

The main idea behind the phenomenology of ballistic transport is the existence of various characteristics lengths.
The first length is the Fermi wave length $\lambda_F=2\pi/k_F$ with $k_F$ the Fermi momentum. Besides the underlying lattice spacing, this is the shortest length of the problem, say typically $\lambda_F \simeq 1-10\, nm$. Typical energies for ballistic transport are usually assumed to be much smaller than the energy scales associated to this length.

Another is the mean free path $\ell_e$, the typical distance an electron travels between successive collisions, say typically $\ell_e\simeq 1-10\, \mu m$. These collisions yield to momentum relaxation by their very nature. We may define the momentum relaxation time $\tau_e$ and write $\ell_e=v_F\tau_e$, where $v_F$ is the Fermi velocity, the electron typical velocity. All collision processes, elastic or not, contribute to the diffusive behavior of the electron gas at large enough scale (at distances larger than $\ell_e$ and times greater that $\tau_e$), with diffusion constant $D_F\simeq v_F\ell_e= v_F^2\tau_e$. When different collision processes are involved, the inverse of the effective relaxation time $\tau_e$ is the sum of the inverse of the relaxation times of each of these processes, $1/\tau_e = \sum_i 1/\tau_i$ (because to a good approximation the collision processes are independent Poisson processes). In other words, as expected, the channel which dominates the early relaxation process is the one with the smallest relaxation time.

We may identify two important length scales related to the relaxation processes. One is that associated to elastic collisions, say on fixed impurities or fixed lattice structure: this modifies the electron momenta ${\bf k'}\not= {\bf k}$ but not their energies $E_{\bf k'}=E_{\bf k}$. They therefore do not induce phase decoherence (they could induce phase shifts via time delays but these are coherent for all electrons). 
The momentum relaxation time associated to elastic collisions, $\tau_\mathrm{elas}$, and hence the associated mean free path $\ell_{\rm elas}$, are largely independent of the temperature, and typically $\ell_{\rm elas}\simeq 1-10\, \mu m$.

Another is the length associated to inelastic collisions, say on phonons. These collisions modify both the momenta and the energy of the electrons, ${\bf k'}\not= {\bf k}$ and $E_{\bf k'}\not=E_{\bf k}$. As a consequence they induce phase decoherence (recall that the phase shift is proportional to the energy shift). The phase decoherence time $\tau_\mathrm{\phi}$ is strongly temperature dependent (because they are highly dependent on the environment behavior, say on the phonon behavior) and hence may be made to be very different from the elastic relaxation time $\tau_\mathrm{elas}$. The phase decoherence length $L_\phi$ depends on the relative scales of $\tau_{e}$ and $\tau_\phi$. If $\tau_{e}\sim\tau_\phi$, then phase decoherence occurs on time scales over which no momentum relaxation has occurred, hence under ballistic electron transport, $L_\phi \simeq v_F \tau_\phi$. On the other hand, if $\tau_{e}\ll\tau_\phi$, then momentum relaxation has occurred between phase-decoherence events, and thus $L_\phi^2 \simeq D_F \tau_\phi$. The phase decoherence length decreases with temperature, typically $L_\phi\simeq 1-10\, nm$ at $T\simeq 300\, K$ while $L_\phi\simeq 1-10\, \mu m$ at $T\simeq 0.1\, K$. 

\begin{figure}\bc
\includegraphics[width=10 cm]{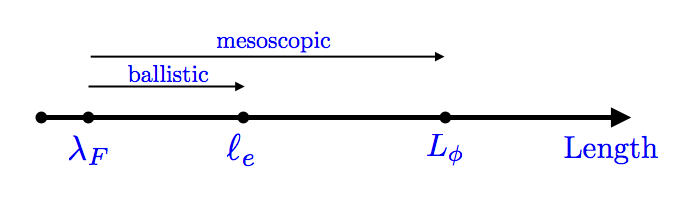}\ec
\caption{\it Length hierarchy in electronic mesoscopic systems: $\lambda_F$ is the Fermi wave length, $\ell_e$ the mean free path, $L_\phi$ the phase decoherence length. }
\label{fig_meso}
\end{figure}

At low enough temperature we enter in a regime in which $\ell_e\ll L_\phi$. In this regime, after the time $\tau_e$, coherence effects are still important. This is the regime of mesoscopic physics, where coherence phenomena may be observed at length scales $L_{\rm obs}$ smaller than the coherence length $L_\phi$. See Figure \ref{fig_meso}.

In this review, we will concentrate on the ballistic regime, at times below $\tau_e$. At critical points of quantum lattice systems, emergent behaviours -- ``quasi-particle'' -- may replace the electrons in the above discussion. In this case, as we will explain in section \ref{sectCFT}, at scales much larger than microscopic length scales and low enough temperatures, quasi-particle may travel ballistically. As temperatures are increased, momentum relaxation occurs because the underlying lattice structure becomes important and breaks collectivity. It is this phenomenon that gives rise to a momentum relaxation time $\tau_e$ in the context of critical systems.

\subsection{Linear response}

The presence of a ballistic or diffusive regime can be detected by studying the response of the electron gas to an electric field.  Within the linear response approximation, the mean electric current ${\bf j}(\omega)$ produced by an electric field ${\bf E}(\omega)$ of frequency $\omega$ is proportional to the electric field, ${\bf j}(\omega)=\sigma(\omega)\, {\bf E}(\omega)$, and the response is coded into the conductivity $\sigma(\omega)$. The response of the electron gas to a constant electric field is encoded into the zero-frequency conductivity ${\rm Re}(\sigma(0))$.

The simplest way of approaching electron transport in metals is the classical Drude model. It consists in describing classically, via Newton's equations, the movement of electrons subject to an electric field ${\bf E}$ and to friction, a force $-{\bf u}/\tau_e$ opposite and proportional to the electron velocity ${\bf u}$. This leads to the so-called Drude formula for the conductivity:
\beq \sigma(\omega)= \frac{\sigma_0}{1- i\tau_e\omega}= \frac{i (n_ee^2/m)}{\omega+i/\tau_e},\eeq
where $\tau_e$ is the friction relaxation time, and the numerator is $\sigma_0={n_ee^2\tau_e}/{m}$ with $n_e$ the electron density and $m$ the electron mass. In particular, ${\rm Re}\, \sigma(\omega) = (\frac{n_ee^2}{m})\frac{\tau_e^{-2}}{\omega^2+\tau_e^{-2}}$.

If $\tau_e$ is finite, the Drude conductivity is regular at zero frequency and is equal to $\sigma_0=(\frac{e^2n_e}{m})\tau_e$. We may notice that the (so-called) Einstein relation relating the conductivity to the diffusion constant is fulfilled, $\sigma_0=e^2\rho_0(\epsilon_F)D_F$, where $\rho_0(\epsilon_F)=n_e/2\epsilon_F$ is the density of states per unit volume. When $\tau_e$ is finite, the electron behavior is clearly diffusive at times larger than the momentum relaxation time $\tau_e$, and distances larger than the mean free path $\ell_e=v_F\tau_e$. On the other hand, if $\tau_e\to \infty$, there is no friction and the electron behavior is, of course, ballistic. In this limit, the real part of the Drude conductivity develops a singular Dirac peak: ${\rm Re}\, \sigma(\omega)= D_0\, \delta(\omega)$ with $D_0={\pi n_ee^2}/{m}$. 

We deduce from this model that ballistic transport is encoded into a singular, delta-function part of the (real part of the) conductivity at zero frequency: we expect that $\sigma(\omega)= D_0\, \delta(\omega) + \sigma_{\rm reg}(\omega)$ (where is $\sigma_{\rm reg}$ regular), if electron transport possesses a ballistic component. The coefficient $D_0$ is called the Drude weight. 

It turns out that this picture survives in the quantum realm. The linear conductivity, which is the linear response of the current under the application of a small electric field, is given by the Green-Kubo formula \cite{Kubo}. This formula may be written in different forms, one of which is the following :
\[ \sigma(\omega)= \frac{1}{L} \int_0^\beta \hskip -.2 truecm d\lambda\int_0^\infty \hskip -.3 truecm dt\, e^{-i\omega t}\, \bra J(0)J(t+i\lambda)\ket_\beta,\]
where $L$ is the size of the sample and $\beta=1/k_BT$ the inverse temperature. Here, $J(t)$ is the total current operator $J(t)=\int \dd x j(x,t)$ where $j(x,t)$ the current density operator at position $x$ and time $t$, and $\bra \cdots\ket_\beta$ denotes the expectation value at thermal equilibrium with temperature $T$. 

The linear conductivity is thus determined by the equilibrium current two-point function $\langle j(0,0) j(x,t)\rangle_\beta$. It is easy to verify from the Green-Kubo formula that the conductivity develops a Dirac peak at zero frequency if the two-point function possesses a component of the form of ballistically traveling waves, $u_+(x-v_Ft) + u_-(x+v_Ft)$. In general, then, we may still decompose the conductivity as
\beq \sigma(\omega)= D_0\, \delta(\omega) + \sigma_{\rm reg}(\omega), \eeq
and non-zero Drude weight $D_0$ signals a ballistic component in the electronic transport. 

The Drude weight can be related to the large-time asymptotic behavior of the current two-point function:
\[ D_0 = \frac{\beta}{2L} \lim_{t\to \infty} \frac{1}{2t}\int_{-t}^t \hskip -.2 truecm ds\, \langle J(0)J(s)\rangle_\beta.\]
This form has been extensively used to bound the Drude weight from below, thus providing proofs of ballistic transport in the linear regime. In particular, when conserved quantities $Q_k$ exist whose second cumulant $\bra Q_k^2\ket_\beta - \bra Q_k\ket_\beta^2$ scale like the volume of the system, the so-called Mazur inequality \cite{Mazur} yields a rigorous estimate for a lower bound on the Drude weight:
\beq\label{drude}
D_0 \geq \frac{\beta}{2L} \sum_k \frac{ \langle JQ_k\rangle_\beta^2}{ \langle Q_kQ_k\rangle_\beta^2}.
\eeq
Here the $Q_k$'s are chosen to be orthonormal $\langle Q_kQ_l\rangle_\beta= \delta_{k,l}\, \langle Q_k^2\rangle_\beta$. This relation has been used to constraint transport properties in simple model systems as well as in integrable quantum spin chain, where novel conserved quantities where introduced for this purpose, see for instance \cite{KSth,Drude_bound}.

\section{General framework} \label{sectgen}

In this review, we mostly concentrate on the properties of quantum steady states (states with flows of energy and charge), instead of those of quantum dynamics. There exist various approaches to theoretically implement  systems out of equilibrium with steady flows. Two main categories are as follows:
\begin{itemize}
\item (Effective reservoirs) One may imagine connecting the system under study to a set of external reservoirs or baths and look for an effective description. The system is then open, in the sense that there are flows of energy or charge between the system and the external reservoir, whose dynamics is effectively described without the full knowledge of the reservoirs themselves. As a consequence the system's dynamics is not unitary, but dissipative. Under the hypothesis that there is no memory effects, or at least that these effects are irrelevant, this effective dynamics is Markovian. It can then be formulated as a semi-group of completely positive maps generated by some Lindblad operator \cite{Lind}. This approach was for instance recently used \cite{Prosen_Review} to described open spin chains in contact with reservoirs at their boundaries.

\item (Hamiltonian reservoirs) One may alternatively englobe the system and the series of reservoirs into a large ``total system''. The dynamics is then unitary as the total system is closed, see for instance \cite{Ruelle}. The Keldysh approach widely used for describing non-equilibrium electronic transport through impurities can be understood within this context, see e.g. \cite{fermion-Keldish}. There, the leads are explicitly described: they are Landau Fermi gases under the approximation that at low energies the spherically symmetric waves dominate (giving rise to one-dimensional free-electron baths). Of course, integrating out the reservoir degrees of freedom, with or without approximations, yields back an effective description as an open system. For instance, this can be performed in the spin-boson models (see e.g. the book \cite{spin-bosons}), in which quantum spins are coupled to an infinite set of harmonic oscillators representing external reservoirs, and, in the weak-coupling limit \cite{DeRoeck}, leads to a Markovian effective dynamics.
\end{itemize}

The most interesting questions relate to phenomena that are largely independent of the way the non-equilibrium state is obtained. Phenomena that are very dependent on the protocol used to put the system out of equilibrium are certainly harder to adequately describe by a theory based on general principles, and most likely are difficult to reproduce experimentally.

The approach taken in this review is a variant of the second one based on unitary evolution of the total system (sometimes referred to as the partitioning approach). We consider the reservoirs to be of the same nature as the quantum system itself. This approach, to our knowledge, was first used in the context of energy transport in classical harmonic chains \cite{rub,spo}, but recently revived interest in it arose thanks to new results in conformal field theory (CFT) \cite{BD2012}. The approach is justified if the phenomena studied are indeed universal enough to be largely independent on the way the system has been driven out of equilibrium; since CFT describes universal emergent behaviours, many of the results and ideas reviewed here are expected to display such independence.

\begin{figure}\bc
\includegraphics[width=15 cm]{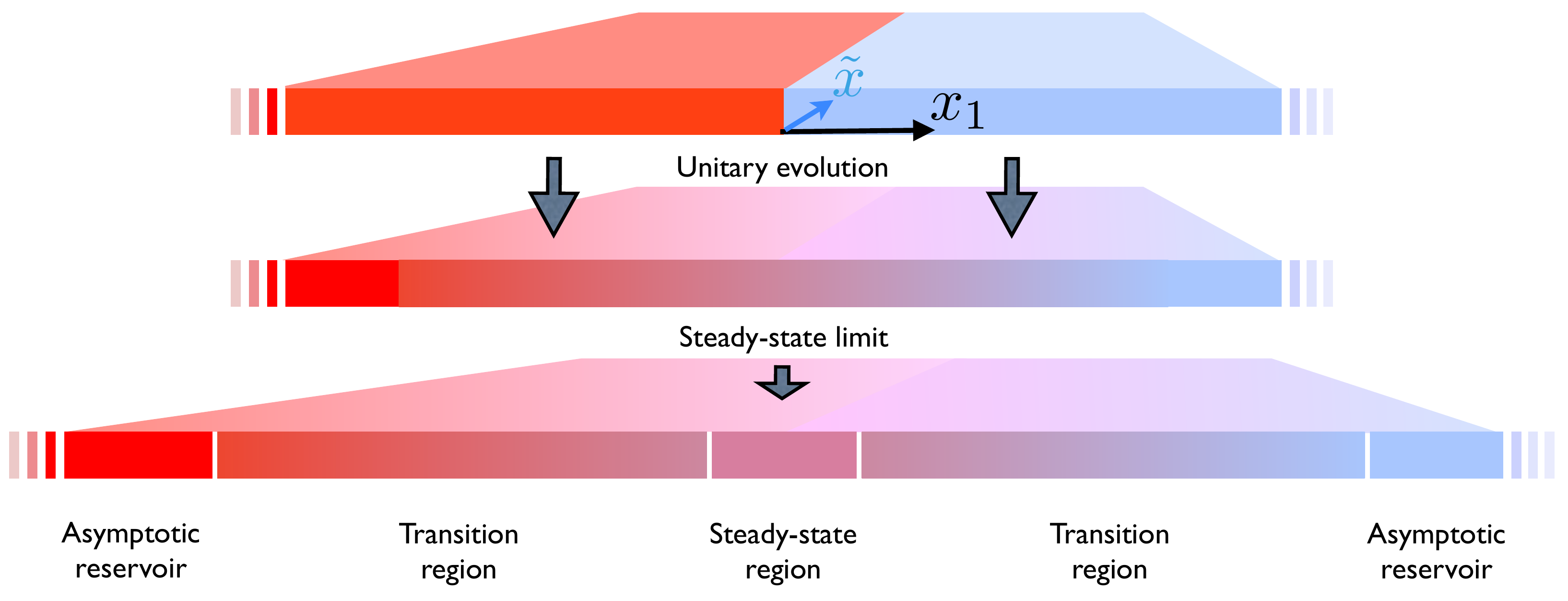}\ec
\caption{\it The partitioning approach. The direction of the flow is $x_1$, and $\t x$ represents the transverse coordinates. After any finite time, asymptotic reservoirs are still present and infinite in length. The central region, around $x_1=0$, is the steady-state region. At very large times, it is expected to be of very large extent.}
\label{fig_steady}
\end{figure}

This approach can be summarized as follows \cite{BD2012} (see Figure \ref{fig_steady}): 
\begin{itemize}
\item[(1)] One prepares independently two semi-infinite homogeneous quantum systems at different temperatures or chemical potentials. In $d$ dimensions, say with coordinates $x_j$, one sub-system extends towards $x_1<0$, while the other towards $x_1>0$, both being infinite in other directions (but other geometries are possible). With the exception of section \ref{sectbeyond}, in this review we concentrate on the case $d=1$.
\item[(2)] At a given time, one puts these systems into contact at the hypersurface $x_1=0$, so that they can exchange energy or charge: immediately after contact has been established, flows of energy or charge are produced. 
\item[(3)] After waiting long enough, relaxation occurs and these flows are expected to reach a steady regime, at least in a domain close enough to the contact region. 
\end{itemize}

The parts of the initial systems far away from the contact region serve as effective reservoirs, providing and absorbing energy and charge as required. The size of the domain in which the steady flows take place may (and usually does) grow indefinitely with time. Yet, for any finite time this domain stays finite, as the information of the connection generically travels at a finite speed (as a consequence, for instance, of the Lieb-Robinson bound \cite{lr}). Since the initial systems were infinitely large, the parts unaffected stay macroscopically large and may indeed behave as reservoirs.

The first question is as to the nature of the steady regime. In generic quantum systems, one would expect thermalization to occur, and thus the final state to be at equilibrium. However, when ballistic components to transport are present, the steady states produced in the neighbourhood of the contact region are out of equilibrium: they carry nontrivial flows. The presence of non-equilibrium flows at large times in this setup provides a definition for non-equilibrium ballistic transport.

We further specify this approach by assuming that the systems, before and after contact, are at or near quantum critical points. They are then described by CFT or perturbations thereof \cite{BD2012}. This naturally has applications to one-dimensional gapless quantum systems and mesoscopic transport, and, as we will see, (near-) critical quantum systems are the most basic examples of systems where ballistic components of transport emerge. At the technical level, making this assumption allows us to maximize the use of the powerful CFT tools, with the potential to enlarge the domain of application of CFTs to out-of-equilibrium situations and to decipher some simple, but non-trivial, examples of non-equilibrium steady states for extended many body quantum systems.

The results reviewed in the manuscript may be understood from two complementary perspectives:
\bi
\item[(i)] CFT out-of-equilibrium may provide information about critical and near-critical systems away from equilibrium, and gives results well beyond linear response. Naturally, the information it gives about critical and near-critical systems depends on how CFT actually emerges at low temperature, and on universality properties of such non-equilibrium states. Although there is no mathematical proof about the emergence of CFT out of equilibrium, there is a bundle of arguments reviewed below, in particular in section \ref{sectCFT}, suggesting that this holds.
\item[(ii)] Independently of its application to criticality and universality, non-equilibrium CFT is by itself of fundamental interest, as it provides a nontrivial framework where new, fully non-linear, non-equilibrium results can be obtained with precision. The number of systems, classical and quantum, for which exact results for non-equilibrium behaviors are known, is very limited. In the classical theory, exact results for specific examples (such as TASEP models) played a crucial role in the formulation of the general macroscopic fluctuation theory for out-of-equilibrium systems, and we expect that non-equilibrium CFT and related subjects may play a similar role in the quantum context.
\ei

\subsection{Ballistic wave propagation in linear response theory}\label{ssectlinresp}

In order to display some of the phenomena that underlie many of the arguments made for quantum critical transport, let us consider the following simple physical situation. Consider a one-dimensional fluid, described by a conserved density ${\tt h}$ and an associated current ${\tt j}$, and assume that ${\tt j}$ is also itself a conserved density, with its own current ${\tt k}$: we have $\p_t {\tt h} + \p_x {\tt j} = 0$,  and $\p_t {\tt j} + \p_x {\tt k}=0$. Relativistic fluids are examples, where ${\tt h}$ is the energy density (averaged over the transverse direction $x_2,\ldots,x_d$ if $d>1$), ${\tt j}$ is the momentum density and ${\tt k}$ is the pressure (in the direction $x_1$). As explained in sections \ref{sectCFT} and subsection \ref{secthydro}, critical and near-critical models generically lead, at least at large scales, to such a fluid description, where ${\tt h}$, ${\tt j}$ and ${\tt k}$ are averages of corresponding quantum observables $h$, $j$ and $k$.

Since ${\tt j}$ is a conserved density, this means that ${\tt J}:=\int \dd x\, {\tt j}(x)$ is a conserved charge. If this fluid can be understood as coming from a quantum system, with quantum conserved charge $J$, then this charge could be used in \eqref{drude}, and since $\bra JJ\ket_\beta>0$, the nonzero Drude weight suggests that there is near-equilibrium ballistic transport. Let us make, instead, an analysis purely based on the fluid picture. At equilibrium, we have ${\tt j}=0$, and the densities ${\tt h}={\tt h}_\beta$ and ${\tt k}={\tt k}_\beta$ take some values, depending on the temperature $\beta^{-1}$. Spanning the temperatures, the ${\tt h}_\beta$-${\tt k}_\beta$ curves gives an equation of state: ${\tt k}_\beta = {\cal F}({\tt h}_\beta)$. Suppose we prepare the two semi-infinite halves of the system at different temperatures $T_l$ and $T_r$, so that ${\tt h}(x,0) = {\tt h}_{\beta_l}\Theta(-x) + {\tt h}_{\beta_r} \Theta(x)$. The full time evolution can be obtained under the assumption that, at all times and at every point, there is local thermalization: the equation of state always holds. If we assume that $T_l\approx T_r$, then to leading order in $\beta_l-\beta_r$ we have ${\tt h}(x,t) = {\tt h}_{\beta} + \delta {\tt h}(x,t)$ and ${\tt j}(x,t)=\delta {\tt j}(x,t)$ where $\beta = (\beta_l+\beta_r)/2$. Combining this with the conservation equations, we get
\beq
	\p_t \,\delta {\tt h} + \p_x \,\delta {\tt j} = 0,\quad \p_t \,\delta {\tt j} + {\cal F}'({\tt h}_\beta)\,
	\p_x \,\delta {\tt h}=0
\eeq
whose general solution is formed out of waves propagating at the ``sound velocity'' $v_s =\sqrt{{\cal F}'({\tt h}_\beta)}$:
\beq
	\delta {\tt h}(x,t) = u_+(x-v_st)+u_-(x+v_st),\quad \delta {\tt j}(x,t) =
	v_s\, \big(u_+(x-v_st)-u_-(x+v_st)\big).
\eeq
Implementing the initial conditions $\delta {\tt h}(x,0) = \delta {\tt h}_{l}\Theta(-x) + \delta {\tt h}_{r} \Theta(x)$ and $\delta {\tt j}(x,0) = 0$, we obtain $2u_+(z) = 2u_-(z) = \delta {\tt h}_{l}\Theta(-z) + \delta {\tt h}_{r} \Theta(z)$. Thus, the solution to the non-equilibrium problem is described by sound waves emitted from the contact point $x=0$ at $t=0$ and traveling in opposite directions, beyond which the fluid is at equilibrium with temperature $T_l$ (for $x<-v_st$) and $T_r$ (for $x>v_st$), and inside which there is a current-carrying non-equilibrium steady state with fluid parameters
\beq\label{jhlinear}
	{\tt j}_{\rm sta} = v_s\frc{{\tt h}_{\beta_l} - {\tt h}_{\beta_r}}{2} =
	\frc{{\tt k}_{\beta_l} - {\tt k}_{\beta_r}}{2v_s},\quad
	{\tt h}_{\rm sta} = \frc{{\tt h}_{\beta_l} + {\tt h}_{\beta_r}}2\qquad (|x|<v_st).
\eeq
Hence we recover near-equilibrium ballistic transport using this fluid picture.

We will see, in the next sections, that the above picture still holds in the fully non-equilibrium steady state of one-dimensional CFT, that it is generically broken (except at critical points) by the presence of conserved charges in integrable models, and that it may be generalized to higher-dimensional CFT and other non-integrable models with ballistic transport.

\subsection{A quantum example: the Heisenberg spin chain}\label{ssectHeis}

Here and for the rest of the review, we set $k_B=\hbar=1$ unless otherwise mentioned.

The fluid arguments in the previous subsection above was a classical, linear-response description of the non-equilibrium setup and of its steady state. In order to illustrate the situation, and what is expected to happen, in a full quantum model, consider instead an anti-ferromagnetic Heisenberg chain of length $2L+2$, with Hamiltonian
\beq\label{Hsum}
	H = \sum_{n=-L}^L h_n,\quad h_n:= \vec\sigma_n\cdot \vec\sigma_{n+1}
\eeq
acting on the Hilbert space ${\cal H}=\prod_{n=-L}^{L+1}(\C^2)_n$. Here $\vec\sigma_n$ is a vector of Pauli matrices acting nontrivially on site $n$ (acting like the identity on other sites).

We divide the chain into two halves: the sites from $-L$ to $0$, and those from $1$ to $L+1$. The associated Hilbert spaces are ${\cal H}_l$ and ${\cal H}_r$, and we have ${\cal H}={\cal H}_l\otimes {\cal H}_r$. One may take the Hamiltonians on each separate half to be
\beqs
	H_l = \sum_{n=-L}^{-1} \vec\sigma_n\cdot \vec\sigma_{n+1},\quad
	H_r = \sum_{n=1}^{L} \vec\sigma_n\cdot \vec\sigma_{n+1}.
\eeqs
Note that $H_l$ and $H_r$ commute with each other: $[H_l,H_r]=0$. The full Hamiltonian has an extra connection between the two halves, so that
\beq\label{HdH}
	H = H_l + H_r + \delta H,\quad \delta H = \vec{\sigma_{0}}\cdot\vec{\sigma_1} = h_0.
\eeq

In the protocol for building the non-equilibrium steady state, the initial state is prepared by thermalizing independently the two halves at temperatures $T_l$ and $T_r$. Hence the initial density matrix is
\beq
	\rho_0 = \frak{n}\lt(e^{-H_l/T_l-H_r/T_r}\rt)
\eeq
where $\frak{n}$ is the normalization map, $\frak{n}(A) = A / \Tr_{\cal H}(A)$. This density matrix factorizes into two commuting density matrices, one for each half of the chain,
\beq
	\rho_0 = \rho_l \otimes \rho_r,\quad
	\rho_{l,r} = \frak{n}\lt(e^{-H_{l,r}/T_{l,r}}\rt).
\eeq
Averages of observables in the initial state are evaluated as usual,
\beq
	\bra \Or\ket_0 := \Tr(\rho_0 \Or).
\eeq

The second step of the protocol is to evolve the state using the full Hamiltonian $H$. This corresponds to evolving the system in time after having suddenly connected the two halves of the chain by adding the single link represented by $\delta H$. The time-evolved density matrix, after time $t$, is $\rho(t) = e^{-\ii Ht}\rho_0 e^{\ii Ht}$. Likewise, the averages of observables in the time-evolved state are
\beq
	\bra\Or\ket_t := \Tr(\rho(t)\Or) = \Tr(\rho_0 \,\Or(t)), \quad \Or(t) = e^{\ii H t}\Or e^{-\ii Ht}.
\eeq

Finally, the third step is to take the steady-state limit. This is the limit where the system size $L$ is very large and the time $t$ is very large, the system size being much larger than the length that can be travelled by excitations in time $t$. That is, $L\gg vt$, where $v$ is a typical propagation speed of excitations (in quantum models, this can be taken as the Lieb-Robinson velocity \cite{lr}; at low temperatures, we may also take it as the Fermi velocity $v_F$ for the excitations). Mathematically, we simply need to take the limits in the following order:
\beq\label{defss}
	\bra \Or\ket_{\sta} := \lim_{t\to\infty} \lim_{L\to\infty}
	\bra \Or\ket_t.
\eeq

We expect the limit to exist whenever $\Or$ is taken to be an observable supported on a finite number of sites (a number of sites that does not depend on $L$ or $t$). This limit describes the steady state, in the sense that it describes the averages of all finitely-supported observables in the steady-state limit. From this, one may wish to define the corresponding steady-state density matrix $\rho_{\rm sta}$, defined on finite systems and whose infinite-$L$ limit reproduces all steady-state averages; it is not guaranteed to exist, but it is a useful concept in many examples.

The first observation is that the steady-state limit may be expected to be ``non-trivial'', in the sense that $\rho(t)\neq \rho_0$ at all times $t$. Indeed, this is because $H$ does not commute with either $H_l$ or $H_r$. Its commutation can be evaluated explicitly:
\begin{eqnarray*}
	&& [H,H_l] = [\delta H,H_l] = [\vec\sigma_0\cdot\vec\sigma_1,\vec\sigma_{-1}\cdot\vec\sigma_0] = 2\ii(\vec\sigma_{-1}\times \vec\sigma_0)\cdot\vec \sigma_{1}\\
	&& [H,H_r] = [\delta H,H_r] = [\vec\sigma_0\cdot\vec\sigma_1,\vec\sigma_{1}\cdot\vec\sigma_2] = 2\ii(\vec\sigma_2\times \vec\sigma_1)\cdot\vec \sigma_{0}
\end{eqnarray*}
This means that $H$ connects $H_l$ to $H_r$, so energy can flow from one half to the other.

A simple observable to concentrate on is the energy current. One way of defining the energy current, flowing from the left to the right, is as the variation with time of the total energy on the right half:
\beq\label{j1}
	j_1 = \frc{\dd}{\dd t} H_r = \ii[H,H_r] = 2(\vec\sigma_0\times \vec\sigma_1)\cdot\vec \sigma_{2}.
\eeq
Of course, we could as well have considered the negative of the variation of the total energy on the left half:
\beq\label{j0}
	j_0 = -\frc{\dd}{\dd t} H_l = -\ii[H,H_l] = 2(\vec\sigma_{-1}\times \vec\sigma_0)\cdot\vec \sigma_{1}.
\eeq
These can be interpreted as the energy currents on site $1$ and $0$, respectively (the boundary site of the right half and the left half, respectively).

This interpretation is further confirmed by another way of defining the energy current (on any site $n$): it is a parity-odd operator $j_n$ supported on a neighborhood of $n$, such that
\beq\label{dhn}
	\frc{\dd}{\dd t} h_n = j_n - j_{n+1}.
\eeq
This is a conservation equation: the time derivative of an operator is the discrete derivative of another operator. It guarantees that $\sum_n h_n$ is conserved in time (up to the details of what happens at the edges of the system $n=-L$ and $n=L+1$). Evaluating the time derivative at $n=0$ for instance, we find $\ii [H,h_0] = \ii [H,\delta H]=\ii[H_l,\delta H]+\ii[H_r,\delta H]$, and we can indeed identity the above operators \eqref{j1} and \eqref{j0}. The result is unique up to addition of a constant site-independent operator, which hence can only be the identity. By the parity-odd condition, this ambiguity is lifted. For general $n$, we then have
\beqs
	j_n=2(\vec\sigma_{n-1}\times \vec\sigma_n)\cdot\vec \sigma_{n+1}.
\eeqs

\begin{figure}\bc
\includegraphics[width=15 cm]{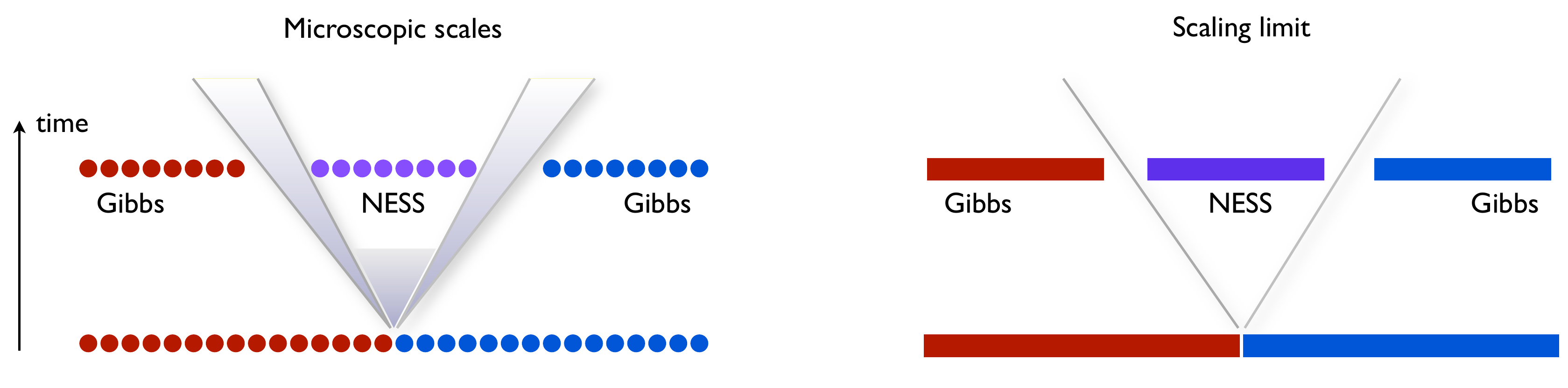}\ec
\caption{\it Shocks and steady state in the non-equilibrium Heisenberg chain. The shocks travel at speeds $v_F$, and have vanishing widths in the scaling limit where observation distances and times are large and temperatures are small compared to microscopic energy scales. The steady state is immediately reached, on universal time scales, after the shocks.}
\label{fig_Heis}
\end{figure}

In the Heisenberg chain, some of the predictions from CFT, valid at low temperatures (that is, in the scaling limit, see subsections \ref{ssectExp} and \ref{ssectcrit}), are then as follows (the derivations are reviewed in sections \ref{sectCFT} and \ref{sectfluctu}):
\bi
\item The steady-state energy current is
\beq\label{jstaheis}
	\bra j_n\ket_\sta = \frc{\pi}{12}(T_l^2-T_r^2) + O(T_{l,r}^3).
\eeq
\item The steady-state energy density is
\beq\label{hstaheis}
	\bra h_n\ket_\sta = \frc\pi{12}(T_l^2+T_r^2) + O(T_{l,r}^3).
\eeq
\item The time-evolution is described by ``shocks'' propagating at the ``velocity of light'' (the Fermi velocity $v_F$) from the contact point. These shocks, as well as the initial transient period before the steady state is established, have widths of microscopic scales, that vanish in the scaling limit, where observation times and distances are large and temperatures are small compared to microscopic energy scales (the energy of a single link, for instance). In the scaling limit, as soon as the shocks are passed, the steady state is immediate, and this steady state is translation invariant. See Figure \ref{fig_Heis}.
\item The full fluctuation spectrum for large-deviations of energy transfer in the steady state is given by the following integrated current cumulants:
\beq\label{fluctuHeis}
	\int_{-\infty}^\infty \dd t_1\cdots \dd t_{n-1}\,
	\bra j_0(t_{n-1})\cdots j_0(t_{1})j_0(0)\ket_{\mathrm{sta}}
	= \frc{\pi n!}{12}\lt( T_l^{n+1} + (-1)^n T_r^{n+1}\rt) + O(T_{l,r}^{n+2}).
\eeq
\ei

We remark that the energy current is greater than 0 for $T_l>T_r$, showing the presence of a non-equilibrium transport of energy from the left to the right halves of the system. Formula \eqref{jstaheis} was verified by numerical evaluations on the Heisenberg chain \cite{Moore12,DeLucaVitiXXZ}. We also remark that Fourier's law is obviously broken: the steady state region is homogeneous, hence there can be no ``temperature gradient'' (no matter how we may wish to define a local temperature), yet there is an energy current\footnote{In the Heisenberg chain, there is no particle transport, hence an energy current is indeed a heat current, the object of Fourier's law.}. Finally, although the Heisenberg chain is integrable, the above is proven in CFT without assuming integrability of the underlying quantum chain: the results are not properties of integrability, but properties of criticality.

\section{Non-equilibrium CFT steady states} \label{sectCFT}

In this section, we develop the basic ideas underlying the low-energy behaviours of critical quantum lattice models, concentrating for simplicity on quantum chains. These are large-scale collective behaviours, whose most important property is universality: they do not depend on the details of the underling lattice structure. Universal critical behaviours are obtained in the scaling limit, which involves large observation distances and times, and low temperatures. The physical theory predicting the emergent, universal behaviours in the scaling limit  is conformal field theory (CFT), and perturbations thereof (giving more generally quantum field theory (QFT)). CFT gives us simple ways of extracting universal information about non-equilibrium steady states. It provides the tools to obtain an explicit construction of the steady state in the scaling limit, as a functional on a large family of observables. We will develop only the aspects of CFT that are needed for the derivation, see the book \cite{CFTbook} for further details.

\subsection{The universal steady-state limit and pre-relaxation}\label{ssectExp}

The principal characteristic of one-dimensional CFT, that is at the basis of non-equilibrium results for energy and charge transport, is chiral separation. Chiral separation is the statement that emergent quanta of energy or charge may only be of two types, a right-moving and a left-moving type, that are independent of each other, and whose dynamics is trivialized as waves with speeds $\pm v_F$ given by the Fermi velocity of the underlying critical model. This is an extremely nontrivial statement, as the underlying model is usually strongly interacting. The statement solely applies to energy and charge transport\footnote{In fact, it applies in general to transport of conserved densities in CFT.} as other observables, such as particle densities, have much more complicated dynamics. Subsection \ref{ssectcrit} is primarily concerned with providing arguments of such chiral factorization, based on fundamental principles of criticality.

Consider the protocol discussed in section \ref{sectgen}, applied to systems described by generic CFTs. That is, much like in subsection \ref{ssectHeis}, we imagine preparing two semi-infinite isomorphic copies of a one-dimensional system whose dynamical properties are described by CFT, one at temperature $T_l$ (on the left) and the other at $T_r$ (on the right). We then put them into contact such that the total system is homogeneous and described by the same CFT.

A non-equilibrium steady state is produced whose properties are induced by the ballistic, wave-like character of transport in CFTs: when the subsystems come into contact, energy or charge propagates at the Fermi velocity as independent waves coming from the left and right reservoirs, and the flow stabilizes in a domain bounded by the propagating wave fronts. At a time $t$ after the contact has been made, this steady-state domain is typically of size $2v_Ft$, see the picture on the right in Figure \ref{fig_Heis}. Clearly, in order to observe steady behaviours, observations must be performed inside this domain. Therefore, the size $\ell_\mathrm{obs}$ of an observable with respect to the connection point -- the distance of the furthest point of the support of the observable to the connection point -- must be smaller than $v_Ft$. As the steady domain should also be smaller than the size $L$ of the total system, the steady-state limit is
\[ L\gg v_Ft \gg \ell_\mathrm{obs}.\]
This translates into the definition of the steady state alluded to in the previous section: for any observable $\mathcal{O}$ of finite size $\ell_\mathrm{obs}$ with respect to the origin, the steady state is defined by the large $L$, and then large $t$ limits, in this order, see \eqref{defss}.

There are thus shock fronts propagating along the light cone, at velocities $\pm v_F$. The shock-wave description indicates that, within the CFT regime and because of its ballistic nature, the steady state is instantaneously attained inside the light cone emerging from the contact point.  Away from the light cone, the system state is a Gibbs state, with temperatures $T_l$ on its left and $T_r$ on its right. These statements are exact in the scaling limit, and as probed by energy observables (right- and left-moving energy densities); other observables will generically not be subject to sharp shock waves.

Let us set $v_F=1$ for simplicity. At thermal equilibrium, the mean energy density ${\tt h}$ of CFT is proportional to the square of the temperature ${\tt h}= \frac{c\pi}{6} T^2$ \cite{Cardy86,Affleck86}, where $c$ is the so-called central charge (it, naively, counts the number of the CFT degrees of freedom). Half of the energy is carried by the left-movers, and the other half by the right-movers. Hence, inside the light cone the energy density is composed of the right-movers coming from the left, which carry a mean energy density $\frac{1}{2} {\tt h}_l= \frac{c\pi}{12} T_l^2$, and the left-movers coming from the right, which carry a mean energy density $\frac{1}{2} {\tt h}_r= \frac{c\pi}{12} T_r^2$. If $T_l\neq T_r$, there is therefore a nonzero mean energy current inside the light cone, and the steady-state energy current and density are
\begin{eqnarray}\label{jsta}
	\bra j\ket_\sta &=& \frac{1}{2}({\tt h}_l-{\tt h}_r) = \frac{c\pi}{12\hbar}k_B^2(T_l^2-T_r^2),\\
	\bra h\ket_\sta &=& \frac{1}{2}({\tt h}_l+{\tt h}_r) = \frac{c\pi}{12\hbar}k_B^2(T_l^2+T_r^2).\label{hsta}
\end{eqnarray}
Here we have temporarily re-instated the constants $\hbar$ and $k_B$, and we note that the last right-hand sides in the above equations are independent of the Fermi velocity $v_F$, thus fully displaying the universality of the result. These results indeed reproduce \eqref{jstaheis} and \eqref{hstaheis}. This is a direct extension of the linear-response picture that led to the formulae \eqref{jhlinear}.

This description is exact within CFT. CFT, of course, only describes the low-energy behaviour of critical systems. In general, there are two sources of departure that may be considered, one leading to an unimportant short-time effect, the other to a more subtle large-time effect.

The first is that at the connection time, a very large amount of energy is injected into the system: the energy of a single site, which is far above the low-energy universal regime. This creates a large amount of non-universal, incoherent processes. As these processes excite high-energy states, they break chiral factorization and the CFT description. However, decoherence means that their effect is fast damped, except for those transported along the edges of the light cone. Therefore, after the corresponding microscopic time scale $\tau_{\rm micro}$, these processes do not contribute to the non-equilibrium energy flow and density inside the light cone, which is dominated by (long-lived) low-energy ballistic quantas coming from the left and right reservoirs. See the pictorial representation on the left in Figure \ref{fig_Heis}.

The second is that, at any nonzero temperatures $T_l$ and $T_r$, there are departures from CFT, which become smaller as the temperatures are made smaller (these departures are thus irrelevant from the viewpoint of the renormalization group). These departures affects the ballistic quantas from the reservoirs. What do we expect, then, for the behaviour of real gapless systems, where such effects are taken into account? At times short enough after the contact has been established, but greater than $\tau_{\rm micro}$, the system's behaviour is still expected to be governed by the ballistic character of the gapless quanta. Thus, after short times, there is a current carrying non-equilibrium steady state inside the light cone, and Gibbs states, at different temperatures, on both sides of the light cone. This is, however, a pre-relaxation state: the medium-energy hamiltonian eigenstates, which are  probed due to the nonzero temperatures, will eventually contribute and break the CFT description, providing for a finite life-time of the quantas. This is a phenomenon that is similar to that of pre-thermalization in the context of the dynamics of weakly non-integrable models \cite{MKpretherm,Bpretherm} (but the use of the term ``thermalization'' suggests equilibrium properties, and hence does not apply well to the present situation). Using the phenomenology of mescoscopic physics, the CFT picture remains valid at times shorter than the ensuing momentum relaxation time $\tau_e$, and as long as the steady-state domain inside the light cone is smaller than the mean free path $\ell_e$.

Thus, the universal steady-state limit, describing ballistic flows carried by low-temperature emergent collective behaviours, must take into account $\tau_e$. The limit is
\beq
	v_F^{-1}L,\,\tau_e \;\gg\; t\; \gg \;v_F^{-1}\ell_\mathrm{obs},\,\tau_{\rm micro}.
\eeq
In order for $\tau_e$ to be made large, it is clear that temperatures must tend to zero, in which case energy current and density vanish. Hence, the precise mathematical limit for the universal steady state must involves a re-scaling of the physical quantities. For an observable $\Or$ with scaling dimension $d_\Or$, the universal steady-state limit is, instead of \eqref{defss},
\beq\label{defsstaue}
	\bra \Or\ket_{\sta}^{\rm universal} :=
	\lim_{t\to\infty} \lim_{L,\tau_e\to\infty}
	\tau_e^{d_\Or} \bra \Or\ket_t.
\eeq
For instance, for the energy density observables in one space dimensions, we have $d_h=d_j = 2$.

\begin{figure}\bc
\includegraphics[width=10 cm]{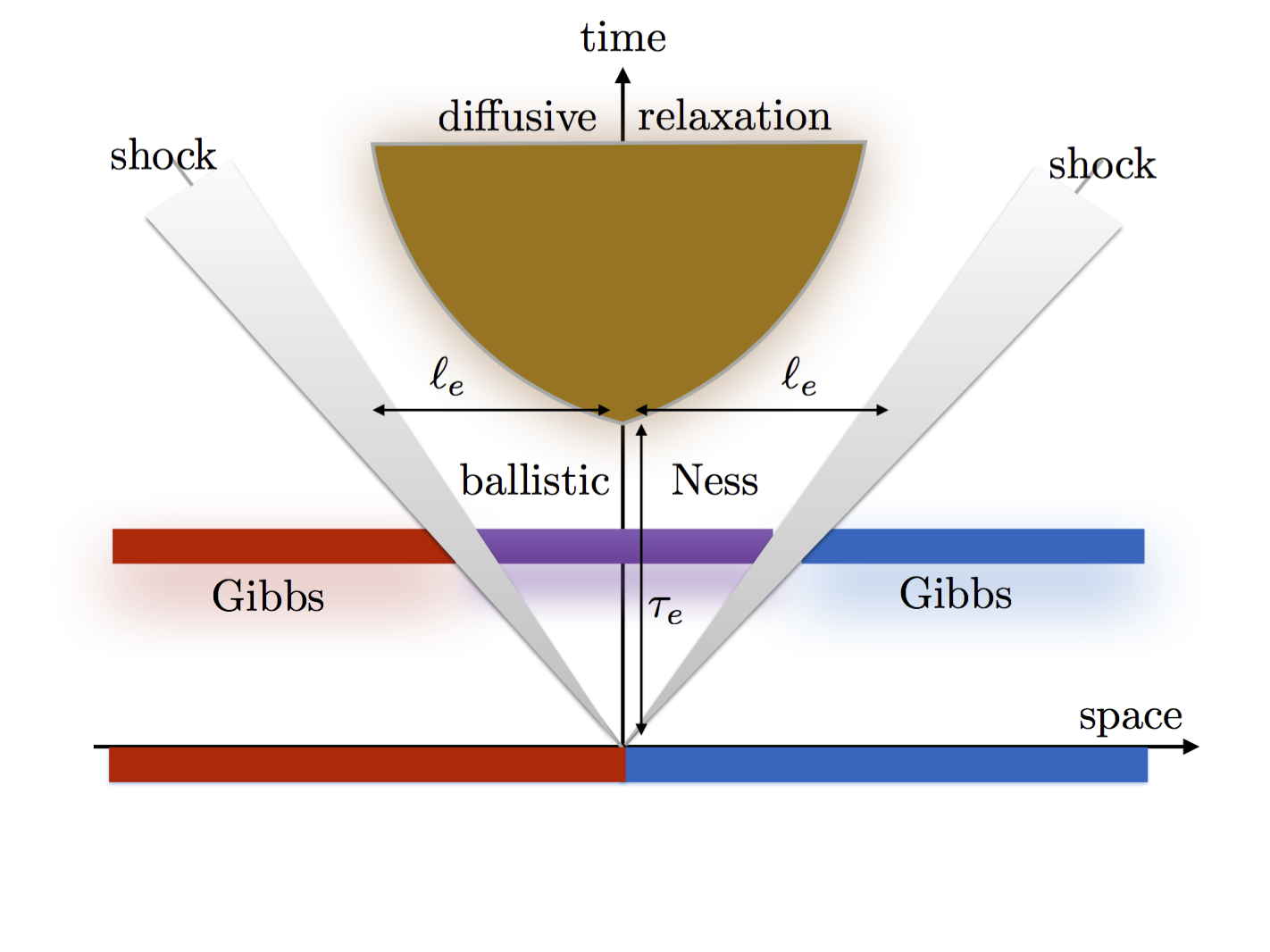}\ec
\vskip -1.5 truecm \caption{\it The universal steady-state regime for energy transport is a ballistic regime situated inside a light cone at intermediate times between the microscopic time scale for pre-relaxation $\tau_{\rm micro}$, during which the extra energy associated to the connection of a single link is dissipated, and the momentum diffusion time $\tau_e$, after which diffusive effects damp the current and generically give rise to equilibration and thermalization. At the edges of the light cone, shocks are present, which should become smoother with time due to diffusive effects.}
\label{fig_crossover}
\end{figure}

For times greater than $\tau_e$, we expect that the shock fronts become smoother as time increases, and that a crossover from ballistic transport to diffusive transport develops inside the light cone domain (see Figure \ref{fig_crossover}). At times much larger then $\tau_e$, full equilibration and thermalization generically occurs.

Some comments are in order:
\bi
\item The light cone may also expand as a consequence of the diffusive processes. However, recall that, even in fully diffusive systems, without ballistic transport, a light cone effect is usually present due to the finite Lieb-Robinson velocity (although there are systems whose Lieb-Robinson velocity is infinite). The light cone might be expected to expand if $v_F$, characteristic of low-energy behaviours, is not equal to the Lieb-Robinson velocity.

\item In the diffusive regime, equilibration and thermalization are expected. However, this expectation is modified in integrable models. In these cases, equilibration may never occur, and steady flows may always exist (in fact even at high temperatures, see subsection \ref{sectbeyondint}), due to the presence of parity-odd conserved quantities that overlap with the energy current or the other currents considered. These conserved quantities will guarantee the stability of ``dressed'' right-moving and left-moving energy quanta, which at low energies smoothly connect with those of CFT. This is the case, for instance, in the Heisenberg model, where the first nontrivial conserved quantity is $J = \sum_n j_n$, the integrated current itself (exactly as in the example of the fluid considered in subsection \ref{ssectlinresp}). In particular, the steady-state limit \eqref{defss}, in the case of the Heisenberg model, is expected to be formally the correct limit, without the modification present in \eqref{defsstaue}. In other integrable models, equilibration may occur if no appropriate parity-odd conserved charges are present that guarantee the stability of dressed chiral quantas, but this equilibration will be to a generalized Gibbs ensemble instead.

\item The shock structure and the presence of the microscopic time scale $\tau_{\rm micro}$ associated to the connection of the single link are both expected to be robust, at low temperatures, to the presence of integrability.

\item The exact results, within non-equilibrium CFT, for energy transport go much beyond the linear hydrodynamics reviewed in subsection \ref{ssectlinresp}: CFT results contain all orders in the dimensionless variable $(T_l-T_r)/(T_l+T_r)$. Similarly, the fluctuation results \eqref{fluctuHeis} derived in section \ref{sectfluctu}, the exact construction of the steady state \eqref{rho_sta}, and the related full correlation results \eqref{corrfctsta}, contain all orders in $(T_l-T_r)/(T_l+T_r)$. 

\item Universality relies on the emergence of CFT at low temperatures (subsection \ref{ssectcrit}). In addition to the general arguments made above, the fact that, both in steady-state averages (see full derivation in subsection \ref{ssectscat}) and in the large-deviation function (section \ref{sectfluctu}), no UV divergencies appear (that is, the same UV regularization is used both in the initial state and in the evolution Hamiltonian), may also be argued to imply universality, as discussed in the literature \cite{BD-long}, even though this does not provide a mathematically rigorous proof. Available exact results in the Ising model \cite{LVBD2013}, where the emergence of CFT out of equilibrium has been demonstrated and deviations from the CFT results may be controlled (including for the large deviation function), further confirm universality. There are also non-trivial verifications, both numerical \cite{Moore12} and experimental \cite{Schwab,Pierre_et_al}, that check universality beyond the linear regime. In particular, note that the experiment of \cite{Pierre_et_al} did verified the $T_{l}^2-T_r^2$ dependence in the mean energy flow (beyond linear response), giving an experimental hint towards universality and emergence of CFT.
\ei

\subsection{Critical points and chiral factorization}\label{ssectcrit}

Quantum lattice models at criticality display very special low-energy behaviours. The main property of criticality is scale invariance: at critical points, physical quantities involving only states of low energies are scale invariant or co-variant. Combined with other physically motivated (and often observed) symmetries, this leads to powerful predictions for these low-energy behaviours.

The assumptions made are that at low temperatures, in thermal states of homogeneous, critical hamiltonians, the following three groups of symmetries emerge: 
\bi
\item[(i)] continuous translation invariance,
\item[(ii)] Lorentz invariance, and
\item[(iii)] scale transformation invariance.
\ei
This emergence is based on the idea that in critical systems, the set of low-energy eigenvalues form, in the thermodynamic limit, a continuum above the ground state without gaps, and energy eigenstates that contribute to the leading behaviours of observables in low-temperature thermal states are those that are very near to the ground state. It is important to emphasize that the energies involve are much smaller even than the energy associated to a single link on the quantum chain.

\noindent (i) {\em Continuous translation invariance.} In homogeneous thermal states on regular quantum lattices, discrete translation invariance occurs naturally: the state is invariant under translations by one lattice site in a symmetry direction of the regular lattice. Let us denote the translation operator by $\tt T$, and concentrate on quantum chains. Then we may organize the energy eigenstates into simultaneous $\tt T$ and $H$ eigenstates, and characterize them by an energy $E$ and a wave number ${ k_1}\in[-\pi,\pi)$, with ${\tt T}^{n}|E,k_1\ket = e^{\ii k_1n}|E,{ k_1}\ket$.

This, however, is not powerful enough to give rise to the full non-trivial behaviours of critical systems. But as mentioned, we are only interested in the low-energy sector, where states have energy very near to that of the ground state. Such low-energy states are describable by a modification of the ground state wave function where additional contributions from the Hamiltonian densities at different points almost cancel each other (so as to give a small energy difference, much smaller than the energy scale associated to a single link on the chain). Intuitively, since the Hamiltonian is assumed homogeneous, we may expect this to happen if the modification is slowly varying, with a wave-like amplitude on large scales. This is in the spirit of single-particle quantum mechanics, where the energy of eigenstates increases with the total number of waves (crests and troughs)\footnote{This points to the powerful re-interpretation of quantum lattice models as models for quantum quasi-particles in interaction, at the basis of the Bethe ansatz.}.

Given such a situation, it is clear that low-energy states must have very small wave number $k_1$. Rescaling as $p_1=k_1\xi$ and $n=x\xi$, we now have a momentum operator, $P$, with eigenvalues $p_1\in\R$, and we may write the effective translation operator, acting on low-energy states, as an exponential of $P$, that is ${\tt T}^{ik_1n} = e^{\ii Px}$.
Every local observable can then be, after appropriate re-scaling, characterized by a continuous position $x$, and we have the infinitesimal action
\beq\label{POr}
	[P,\Or(x)] = \ii\p_x \Or(x).
\eeq

Using translation invariance, we may re-write the Hamiltonian as an integration over a continuous density, instead of a sum over lattice sites as in \eqref{Hsum},
\beq\label{Hint}
	H = \int \dd x\,h(x).
\eeq
Also, by homogeneity, we expect that it be possible to write $P$ as an integral over positions,
\beq\label{Pp}
	P= \int \dd x\, p(x),
\eeq
of some observable $p(x) = e^{-\ii P x}\,p(0)\, e^{\ii Px}$. By virtue of \eqref{POr}, it must be possible to choose $p(x)$ so that it be local: it is the local operator whose commutator with $\Or(x')$ is supported at $x=x'$ with weight $\ii\p_x \Or(x)$.

Since translation is a symmetry, we must have
\beq\label{HP}
	[H,P] = 0.
\eeq
Using \eqref{Pp}, this implies that $[H,p(x)] = \p_x(\cdots) $
where $\cdots$ is some local observable\footnote{We skip technical details. The main argument is that the conservation equation \eqref{HP} holds in the sense of density matrices with insertions of local observables, and local observables at infinity drop out in this context.}. That is, there must exists a local observable $k(x)$ such that
\beq\label{pk}
	\p_t p + \p_x k = 0.
\eeq
Finally, the continuous version of \eqref{dhn} is obviously
\beq\label{hj}
	\p_t h + \p_x j = 0.
\eeq

Therefore, from continuous translation invariance near a critical point we have obtained the existence of a hamiltonian $H$ and a momentum operator $P$, of densities $h(x)$ and $p(x)$, and of the associated currents $j(x)$ and $k(x)$ with the conservation equations \eqref{hj} and \eqref{pk}.

Continuous translation invariance is, in general, crucial in order to have emergent ballistic behavior. Indeed, the lattice structure is expected to lead to back-scattering effects that will induce diffusion and decay of the energy current. Hence, in general, only at low energy may we expect to have, in critical models, an emerging non-equilibrium energy current. As mentioned, in integrable models, such as the Heisenberg chain, the situation is often different because of the presence, at all energy scales, of extra conservation laws.

\noindent (ii) {\em Lorentz invariance.} The emergence of Lorentz invariance is a rather non-trivial statement. A necessary condition is that, at large scales (that is, at the scale of the collective waves forming the low-energy eigenstates), both space and time must scale in the same way -- for otherwise, any maximum velocity could naturally  be scaled away. We say that the dynamical exponent $z$, in $x\sim t^z$, is equal to one. With unit dynamical exponent, it might then be acceptable that low-energy waves cannot propagate faster than a certain speed (as mentioned, the existence of a maximal speed, at all energies, is a consequence of  the Lieb-Robinson bound \cite{lr}). With translation invariance and a maximal propagation speed, one may then argue (as Einstein did) for Lorentz invariance.

Lorentz invariance may be obtained by requiring that the boost operator
\beq\label{Bh}
	B = -\ii\int \dd x\,xh(x),
\eeq
which generates infinitesimal Lorentz transformations $x\p_t$ (here at $t=0$), satisfies the relations
\beq
	[B,H]=P,\quad [B,P]=H.
\eeq
The second relation is a direct consequence of \eqref{POr}, so does not impose any condition. The first relation, in combination with \eqref{hj}, then implies $P=\int \dd x\,j(x)$, and thus, since this is an operator relation, leads to the equality between the energy current and the momentum density (up to total derivative terms that can be absorbed into appropriate redefinitions):
\beq\label{jp}
	j=p.
\eeq

\noindent (iii) {\em Scale invariance.} We may take a similar approach as above: we require that the operator
\beq
	D = -\ii \int \dd x\,xp(x),
\eeq
which generates infinitesimal scale transformations $x \p_x$ (again, here at $t=0$), satisfies
\beq
	[D,H] = H,\quad [D,P] = P.
\eeq
Again, the second relation is immediate, and the first implies
\beq\label{kh}
	k=h.
\eeq

{\em Combining.} We now combine \eqref{hj}, \eqref{pk}, \eqref{jp} and \eqref{kh}. The result is strikingly simple: the two conservation equations
\beq
	\p_t h + \p_x p = 0,\quad \p_t p + \p_x h = 0
\eeq
can be combined into ``chiral factorization'':
\beq
	\b\p T = 0,\quad \p \b T = 0
\eeq
where
\beq\label{Thp}
	T = \frc{h+ p}2,\quad
	\b T = \frc{h- p}2,\quad \p = \frc12 (\p_x-\p_t),\quad \b\p
	= \frc12 (\p_x + \p_t).
\eeq
Thus the quantum time evolution is
\beq\label{evol}
	T(x,t) = T(x - t),\quad
	\b T(x,t) = \b T(x + t)
\eeq
(where here and below, a single argument means that the operator is evaluated at time $t=0$). As a consequence of \eqref{evol}, in any state $\bra\cdots\ket$ that is clustering at large distances, and invariant under time evolution, factorization occurs:
\beq\label{chiralfact}
	\bra\prod_i T(x_i) \prod_j \b T(y_j)\ket
	= \bra \prod_i T(x_i)\ket\bra \prod_j \b T(y_j)\ket.
\eeq
Indeed, we can just evolve in time the left-hand side for an infinite time and use invariance and clustering to obtain the right-hand side. This is chiral factorization.

\subsection{The steady state from a scattering formalism and the light-cone effect}\label{ssectscat}

A fundamental concept for the study of non-equilibrium steady states is that of the McLennan-Zubarev (MZ) non-equilibrium statistical operator \cite{MZoperator}. This is an operator constructed using the initial non-homogeneous densities and the time-integral of the associated non-equilibrium currents, which allows to represent steady states and other non-equilibrium states via statistical ensembles generalizing Gibbs ensembles. In the context the partitioning approach in homogeneous quantum systems, as far as we are aware, ref. \cite{Tasaki} was the first to derive an explicit, rigorous expression for a quantum steady state (in a free-fermion hopping model), showing its form as an MZ ensemble. In this example, the MZ operator was explicitly represented using scattering states arising from the $C^\star$-algebraic formulation of quantum chains. Independently, Ruelle \cite{Ruelle} (see also Bratteli and Robinson \cite{BR1}) developed the idea of a $C^\star$-algebraic scattering formulation of steady states (and other non-equilibrium states) in much more generality. Various authors \cite{steady-rigorous} have then rigorously constructed non-equilibrium steady states in integrable quantum chains with free-fermion representations, see the reviews \cite{steady-rev}. In the context of massive quantum field theory, this general idea was used in order to formally describe the steady-state density matrix \cite{doyqft}. Non-equilibrium density matrices (or operators) were also studied in many related works, e.g. on Luttinger liquids \cite{Protopopov,Mintchev} and impurity models \cite{impurdens,DA2006}. In all cases, the result is physically intuitive: scattering states representing flows of quantas coming from the left (right) are distributed according to the thermal distribution of the left (right) reservoir. See subsection \ref{sectbeyondint} where the idea is reviewed in homogeneous free-particle models. In the present subsection, we show how the tools of CFT allow us to obtain similar results for CFT non-equilibrium steady states. We review the results derived in \cite{BD2012}, which broadly followed the approach proposed by Ruelle \cite{Ruelle}. Applications of this approach has been further developed in \cite{BD-long,LVBD2013,BDM2015,DeLucaVitiXXZ}. Inhomogeneous quantum quenches have also been considered in \cite{Cardy-Sotiriadis}.

We now use CFT tools in order to construct non-equilibrium steady states within the protocol described in section \ref{sectgen}. Recall that for any local observable $\mathcal{O}$ in CFT with finite support, the steady state is defined by:
\beqs
	\bra \mathcal{O} \ket_{\sta} := \lim_{t\to\infty} \lim_{L\to\infty}
	\bra \mathcal{O} \ket_t
\eeqs
with 
\beqs
	\bra\Or\ket_t := \Tr(\rho_0 \,e^{\ii H t}\Or e^{-\ii Ht}).
\eeqs
Here $\rho_0=\rho_l\otimes\rho_r$ is the initial density matrix and $H$ the hamiltonian for the coupled system; this is the hamiltonian coding for the time evolution once the two halves of the system have been coupled through the point contact.

Let $H_0= H_l+H_r$ be the hamiltonian for the two uncoupled systems, where $H_{l,r}$ are the hamiltonians for the left/right parts of the system. The initial density matrix $\rho_0$ is steady with respect to $H_0$, so that it commutes with the uncoupled evolution operator $e^{-\ii tH_0}$. Hence we have $\rho_0= e^{+\ii tH_0}\rho_0 e^{-\ii tH_0}$, or alternatively
\[  \Tr(\rho_0 \,e^{+\ii H t}\Or e^{-\ii Ht}) =  \Tr(\rho_0\,e^{-\ii tH_0} \,e^{+\ii H t}\Or e^{-\ii Ht}\, e^{+\ii tH_0}).\]
The operators $e^{\pm \ii tH}$ or $e^{\pm \ii tH_0}$ do not converge as $t\to\infty$ because of oscillating phases, but the products $e^{\mp \ii Ht}\, e^{+\pm\ii tH_0}$ may converge -- because the phases compensate. Let us define the $S$-matrix by $S:= \lim_{t\to\infty} \lim_{L\to\infty} 
e^{+ \ii H_0t}\, e^{-\ii tH}$. Its precise definition is through its action on local operators:
\beq
S(\mathcal{O}):= \lim_{t\to\infty} \lim_{L\to\infty}\, e^{-\ii tH_0} \,e^{+\ii H t}\,\Or\, e^{-\ii Ht}\, e^{+\ii tH_0}.
\eeq
The steady state may then be represented as
\beq\label{Ssta}
\bra \mathcal{O} \ket_{\sta} = \bra S(\mathcal{O}) \ket_0
\eeq
where $\bra \cdots \ket_0$ denotes the expectation with respect to the initial state $\rho_0$.
This formula has a simple interpretation: the $S$-matrix codes for the dynamics with respect to which the state is steady. This dynamics possesses many steady states and the initial state selects the actual steady state, encoding the appropriate asymptotic-reservoir conditions.

The $S$-matrix also possesses a simple interpretation: its action amounts to first evolving the operator forward in time with the coupled hamiltonian $H$ for a long time period, and then to evolve this operator backward in time with the uncoupled hamiltonian $H_0$, for the same time duration.

This is easily implemented in CFT thanks to chiral factorization. The evolution with $H$ is that already described in \eqref{evol}. The evolution with $H_0$ receives simple modifications due to the boundary at $x=0$: since $H_0$ has, for all $x\neq0$, the same density as $H$,  the same conservation equations \eqref{hj}, \eqref{pk}, \eqref{jp} and \eqref{kh} hold and thus the evolution with $H_0$ of local operators away from the origin is still given by \eqref{evol}. However, at the origin, one must implement the condition that the energy may not flow through it (the energy current is zero) \cite{Cardy_boundary}:
\beq
	T(0) = \b T(0)\qquad \mbox{(under $H_0$)}.
\eeq
Therefore, under evolution with $H_0$, when a local operator hits the origin one simply exchanges $T\leftrightarrow \b T$ in order to continue the evolution without crossing it: from the viewpoint of local right- and left-moving energy densities, there is a simple reflection.

Let us look at the $S$-matrix action on the stress tensor components $T(x)$ and $\bar T(x)$. Consider $S(T(x))$ for $x<0$. Since $T$ describes right movers, the forward evolution with the coupled dynamics moves $T(x)$ to the left at the point $x-t$, far away from the origin for $t$ very large, while the backward evolution with the uncoupled dynamics then moves this operator back to its initial position because the origin is never encountered. Hence $S(T(x))=T(x)$ for $x<0$. Let us now look at $T(x)$ but with $x>0$. The forward evolution with the coupled dynamics still moves this operator far to the left at point $x-t$. For $t>|x|$, this operator crosses the origin during its displacement, but this crossing is allowed by the coupled dynamics, as the coupled system is homogeneous. The backward evolution with the uncoupled dynamics however does not allow crossing back, as the system is split and the operators are reflected at the boundary. Therefore the operator is not brought back to its original position, but rather to the position symmetrically opposite with respect to the origin. During the reflection process, the operator $T$ and $\bar T$ are exchanged. Hence $S(T(x))=\bar T(-x)$ for $x>0$. In summary we have:
\begin{eqnarray}
S(T(x))&=& T(x),\quad ~~\mathrm{for}\ x<0,\\
S(T(x))&=& \bar T(-x),\quad \mathrm{for}\ x>0.
\end{eqnarray}
Similarly,
\begin{eqnarray}
S(\bar T(x))&=& T(-x),\quad \mathrm{for}\ x<0,\\
S(\bar T(x))&=& \bar T(x),\quad ~~~\mathrm{for}\ x>0.
\end{eqnarray}
This determines the steady state on any product of the stress tensor components, that is, on any operators of the form $\prod_k T(x_k)\prod_l\bar T(y_l)$. This is clearly the simplest set of operators we can look at but this is also all what we need if we aim at describing the energy flow. 

We must then evaluate the right-hand side of \eqref{Ssta}. For this, one may use the unfolding principle for evaluating correlation functions in the energy sector in boundary CFT. The argument is as follows. Far from the boundary, one expects a thermal bulk state where the boundary is not felt anymore\footnote{Such a clustering was shown mathematically, in particular for finite-temperature quantum chains with finite local degrees of freedom, in \cite{KGKRE}.}, $\lim_{x\to\mp\infty}\bra \prod_i \Or(x+x_i)\ket_0 = \bra \Or\ket_{\beta_{l,r}}$. Also, recall that $\bra\cdots\ket_0$ is invariant under the $H_0$ evolution, $\bra \prod_i \Or(x+x_i)\ket_0 =  \lim_{t\to\infty} \bra e^{\ii H_0 t}\prod_i \Or(x+x_i) e^{-\ii H_0 t}\ket_0$. Using the simple boundary-reflection evolution, this simplifies the evaluation of averages $\bra\cdots\ket_0$: they are products of bulk averages, where both sides are independently ``unfolded'' ($\b T(x)\mapsto T(-x)$) so as to put right-movers and left-movers that were at positions $x<0$ onto a single line at temperature $\beta_l$, and those that were at $x>0$ onto another, independent line at temperature $\beta_r$.

Using the CFT thermal bulk average formulae \cite{Cardy86,Affleck86}
\beqs
	\bra T\ket_\beta = \bra \b T\ket_\beta = \frc{c\pi}{12} T_l^2,
\eeqs
with $c$ the central charge, the steady-state results \eqref{jsta} and \eqref{hsta} then immediately follow from the above discussion, for instance with $x<0$
\beqa
	\bra j(x)\ket_{\rm sta} &=& \bra S(T(x)) - S(\b T(x))\ket_0
	= \bra T(x)\ket_0 - \bra T(-x)\ket_0 \nonumber \\
	&=& \bra T\ket_{\beta_l} - \bra T\ket_{\beta_r}
	= \frc{c\pi}{12} (T_l^2- T_r^2).
\eeqa

We note that the steady-state current takes the form
\beq\label{jlr}
	\bra j\ket_{\rm sta} = \mathfrak{J}(T_l)-\mathfrak{J}(T_r).
\eeq
This is equivalent to the ``additivity property'' $\bra j\ket_{\rm sta}|_{T_1,T_2} + \bra j\ket_{\rm sta}|_{T_2,T_3} + \bra j\ket_{\rm sta}|_{T_3,T_1} = 0$. As noticed in \cite{Moore12}, the relation \eqref{jlr} in general implies that the non-equilibrium current can be obtained purely from the linear-response conductivity $G(T) = \dd\bra j\ket_{\rm sta}/\dd T_l\big|_{T_l=T_r=T} =  \dd \mathfrak{J}(T)/\dd T$ as
\beq\label{Moorerel}
	\bra j\ket_{\rm sta} = \int_{T_r}^{T_l}\dd T\,G(T).
\eeq

Notice that the derivation we just described, considered at finite times, also proves that observables are steady as soon as they enter the light cone, and that the state is a translation invariant bulk state (there are no boundary effects) inside the light cone. Hence, the non-equilibrium steady state is localized inside the light cone, and Gibbs states (where the boundary is still felt) are outside the light cone, with shock waves separating these regions.

We may in fact use the previous construction to identify directly the steady-state density matrix. The noticeable fact \cite{BD2012,Nat-Phys} is that this density matrix is simply a `boosted version' of a CFT density matrix. 
By duality the $S$-matrix acts on density matrices via $\Tr(S^*(\rho)\mathcal{O}):=\Tr(\rho S(\mathcal{O}))$ for any operator $\mathcal{O}$. With this definition, the steady state reads
\[ \rho_\mathrm{sta} = S^*(\rho_0).\]
Recall that  the initial density matrix $\rho_0$ is the Gibbs state $\rho_0= \mathfrak{n}\big(e^{-\beta_l H_l- \beta_r H_r}\big)$ with $H_l=\int_{-\infty}^0 \hskip -0.1 truecm dx(T(x)+\bar T(x))$ and $H_l=\int_0^{+\infty} \hskip -0.1 truecm dx(T(x)+\bar T(x))$ the left and right hamiltonians. To compute $S^*(\rho_0)$ we need to know how $S^*$ acts on the stress tensor components. Using the above action of the $S$-matrix on $T$ and $\bar T$, it is easy to verify that
\begin{eqnarray*}
S^*(T(x))&=& T(x),\quad ~~\, \mathrm{for}\ x<0,\\
S^*(\bar T(x))&=& T(-x),\quad \mathrm{for}\ x<0,
\end{eqnarray*}
and similarly,
\begin{eqnarray*}
S^*(T(x))&=& \bar T(-x),\quad \mathrm{for}\ x>0,\\
S^*(\bar T(x))&=& \bar T(x),\quad ~~\, \mathrm{for}\ x>0.
\end{eqnarray*}
As a consequence, $S^*(H_l)= \int_{-\infty}^{+\infty} \hskip -0.1 truecm \dd x\, T(x)$ and $S^*(H_r)= \int_{-\infty}^{+\infty} \hskip -0.1 truecm \dd x\,\bar T(x)$, and the steady-state density matrix reads
\begin{eqnarray}\label{rho_sta}
\rho_\mathrm{sta} = \mathfrak{n}\big( e^{- \frac{1}{2}(\beta_l+\beta_r) H - \frac{1}{2}(\beta_l-\beta_r) P}\big),
\end{eqnarray}
where $H=\int_{-\infty}^{+\infty} \hskip -0.1 truecm \dd x(T(x)+\bar T(x))$ and $P=\int_{-\infty}^{+\infty} \hskip -0.1 truecm \dd x(T(x)-\bar T(x))$ are the hamiltonian and moment operators for the total system. This is indeed a boosted density matrix with velocity $\frac{\beta_l-\beta_r}{\beta_l+\beta_r}$ and rest-frame temperature $\sqrt{T_l T_r}$. It is important to note that the stationary density matrix is the exponential of local operators, which have strong consequences on spatial correlations (see the next subsection).

This may also be written in its factorized form on the left and right movers,
\beq\label{rho_stalr}
	\rho_\mathrm{sta} = \mathfrak{n}\big( e^{-\beta_l H_+-\beta_r H_-}\big),
\eeq
with $H_+ = \int_{-\infty}^\infty \dd x\,T(x)$ and $H_- = \int_{-\infty}^\infty \dd x\,\b T(x)$, as well as in the form of a MZ operator,
\[
	\rho_\mathrm{sta} = \frak{n}\big(e^{\big[
	-\beta_l H_l -\beta_r H_r - (\beta_l-\beta_r)\int_{-\infty}^0
	\dd t\, j(0,t)}
	\big]
	\big).
\]

Finally, we note that the non-vanishing of the mean energy current implies that the state $\bra\cdots\ket_{\rm sta}$ breaks time-reversal invariance, because the momentum operator is odd under time reversal. This steady state is thus out-of-equilibrium. It also sustains a positive entropy production. Indeed according to standard thermodynamical arguments, the entropy production per unit of time is $\sigma:={T_r}^{-1}(\frac{dE^r}{dt})+{T_l}^{-1}(\frac{dE^l}{dt})$ with $E^{l,r}$ the mean energies of the left/right sub-systems which play the role of reservoirs and $T_{l,r}$ their respective temperatures. By energy conservation, $\frac{dE^r}{dt}=-\frac{dE^l}{dt}=\bra j\ket_{\rm sta}$. Hence the entropy change per unit of time is $\sigma=({T_r}^{-1}-{T_l}^{-1})\, \bra j\ket_{\rm sta}$, which is positive since $\bra j\ket_{\rm sta}\propto (T_l^2-T_r^2)$.

\subsection{Correlations}

Of course, the exact steady-state density matrix allows us to evaluate not only averages, but also all correlation functions of the energy density and current (as well as those of all other operators obtained through their operator product expansions). We may use the density matrix in the form \eqref{rho_stalr}, and recall the general chiral factorization identity \eqref{chiralfact}. In the steady state, all right-movers are thermalized at temperature $T_l$, and all left-movers at temperature $T_r$. Given the linear relations \eqref{Thp}, we can therefore re-write all steady-state correlation functions of energy densities and currents in terms of thermal ones,
\beqa
	\lefteqn{
	\lt\bra \prod_{x\in X} h(x) \prod_{y\in Y} p(y)
	\rt\ket_\mathrm{sta}} &&\n
	&=& \frc1{2^{|X\cup Y|}} \sum_{\cup_i X_i = X\atop \cup_i Y_i= Y}
	(-1)^{|X_4\cup Y_3|}
	\lt\bra
	\prod_{x\in X_1} h(x)
	\prod_{x\in X_2} p(x)
	\prod_{y\in Y_1}h(y)
	\prod_{y\in Y_2}p(y)
	\rt\ket_{\beta_l} \times \n &&
	\qquad \qquad \qquad\qquad\lt\bra
	\prod_{x\in X_3} h(x)
	\prod_{x\in X_4} p(x)
	\prod_{y\in Y_3}h(y)
	\prod_{y\in Y_4}p(y)
	\rt\ket_{\beta_r}\label{corrfctsta}
\eeqa
where the sum is over all partitions $X_1,\ldots X_4$ of $X$, and all partitions $Y_1,\ldots,Y_4$ of $Y$. That is, the steady-state correlation functions of these observables are linear combinations of products of thermal correlation functions at temperatures $T_l$ and $T_r$. This implies, in particular, that steady-state correlation functions cluster exponentially at large distances: correlations of energy currents and densities decay exponentially in energy-carrying steady states of one-dimensional quantum critical systems.

In the full time evolution, these energy-sector correlation functions also feel the sharp shock fronts as in the picture on the right in Figure \ref{fig_Heis}: as soon as all space-time points $(x,t)$ lie within the light cone, the above result is exact. This is true of all correlations of symmetry currents as well (see section \ref{sectcharge} for a discussion of $u(1)$-symmetry currents). As mentioned, other observables, however, are not in general subject to a sharp-shock-front phenomenon: they will rather display algebraic transition regions inside the light cone, with exponents controlled by those of boundary conformal field theory. See for instance \cite{Protopopov} for results concerning correlations of other observables in the non-equilibrium Luttinger model.

One observable of particular interest in the context of quantum models is the entanglement between parts of the system; for instance, entanglement between the left and the right halves is expected to grow as the system starts to evolve, as the two initially independent parts become entangled due to their interaction. Interestingly, beyond this simple expectation, it was found, numerically \cite{EZ} and theoretically \cite{HD15}, that in the harmonic chain, the entanglement negativity, measuring the entanglement between finite regions of the reservoirs, does show both sharp-shock-front behaviours and chiral factorization effects. It additively separates between entanglement contributions from left- and right-movers, and shows sharp changes as the shock fronts are crossed by the regions boundaries. One observes that the sharp change as the regions are ``absorbed'' by shocks is a sudden decrease: this indicates that the entanglement between parts of the system that are within the shocks is large, presumably as an effect of coherent particle pairs emitted at the contact point. The evolution of entanglement in related non-equilibrium contexts, such as quantum quenches, has been widely studied \cite{CCEEqu}, see the reviews \cite{CCreview,EFreview} in the present volume.

\section{Large deviation theory for critical energy transport}\label{sectfluctu}

The modern theory of non-equilibrium thermodynamics puts a strong emphasis on the fluctuation properties of thermodynamic variables, see e.g. \cite{Derrida-revue, MFT,Touch} and references therein. In  thermodynamics, averages over all degrees of freedom of extensive variables are random variables whose probability distribution peaks around an almost-sure value in the thermodynamic limit. This phenomenon of self-averaging means that the average energy, for instance, takes a single value almost surely in the limit of large system sizes: thermodynamic variables are non-fluctuating. For large but finite system sizes, the probability for deviation from this almost-sure value is small. The theory of large deviations is concerned with the way this probability distribution tends to a peak function. It turns out, as was observed a long time ago, that the way it peaks, formalized into the large-deviation function, encodes all of thermodynamics (for instance, the entropy is naturally defined in this way). These ideas are readily generalizable to non-equilibrium systems, where both large-volume and large-time limits may be taken. For instance, general symmetry relations for large-deviation functions, such as the Cohen-Gallavotti fluctuation relations \cite{fluctu} and other generalisation \cite{Jarz}, may be obtained, encoding some of the physics of non-equilibrium systems.

These ideas have been greatly developed for classical systems and lead for instance to the development of the macroscopic fluctuation theory, see \cite{MFT, Bodineau-Derrida}. In quantum theory, however, subtleties arise because of the quantum, in addition to statistical, nature of fluctuations, and because states and observed values are nontrivially related through the measurement processes of quantum mechanics, see e.g. \cite{Espo}.

Consider the statistics of energy transfer in the present setup. In line with the idea of analyzing distribution functions of extensive variables, a natural quantity to consider is the total energy transferred $E(t)$ over a large period of time $t$. In order to properly define this quantity in quantum mechanics, we must define how the values of $E(t)$ are obtained within a quantum measurement protocol.

There are (at least) two natural ways of defining $E(t)$. One is by defining it via indirect measurements of the local energy current $p(0,t)$ over a long period of time. Indirect measurements must be taken, for otherwise, because of a quantum Zeno-like effect, one would expect direct quantum measurements to ``freeze'' the current to a certain value (notwithstanding the potential problem of experimentally implementing a direct local current measurement). Another way, that also avoids the quantum Zeno effect, is by performing two measurements of half the energy difference $\Delta H = (H_r-H_l)/2$ between the left and right halves of the system, one at time $t_0$ (obtaining the value $e_0$) the other at time $t_0+t$ (obtaining the value $e_t$), defining the random variable $E(t) = e_t-e_0$, the change of their observed values over the interval $t$. Note that in all measurement protocols, the initial time $t_0$ at which the protocol starts is also a free parameter.

In the following, we will discuss only the two-measurement protocol, with $t_0$ chosen to be the time at which the two halves of the system are brought into contact, which we set at $t_0=0$ (and briefly comment on other choices in section \ref{sectcharge})

\subsection{Scaled cumulant generating function for two-time measurement protocol}\label{ssectSCGF}

The two-time measurement protocol, with $t_0=0$ the connection time, can be described mathematically as follows. First, the system is in the density matrix $\rho_0$. A measurement is made of the quantity $\Delta H$, whose associated projection operator onto its eigenvalue $e$ we denote by $P_e$. The probability of observing the value $e_0$ is $\Tr(P_{e_0}\rho_0 P_{e_0})$. After this measurement, the density matrix is $\t\rho_0 = \frak{n}(P_{e_0} \rho_0 P_{e_0})= \frak{n}(P_{e_0} \rho_0)$ where in the last equality we used the fact that $\Delta H$ commutes with $\rho_0$. The system is then evolved for a time $t$, and another measurement is made, where the value observed is $e_t$. Given that $e_0$ was observed, the probability of then observing $e_t$ is $\Tr(P_{e_t} e^{-\ii Ht} \t \rho_0 e^{\ii Ht}P_{e_t}) = \Tr(e^{\ii H t} P_{e_t} e^{-\ii Ht} \t \rho_0)$. Combining, the joint probability is
\beq
	\mathbb{P}(e_0,e_t) = \Tr\lt(e^{\ii H t} P_{e_t} e^{-\ii H t} P_{e_0}\rho_0\rt).
\eeq

We now consider the generating function $F(\lambda,t)$ for the cumulants of the differences $E(t) = e_t-e_0$,
\beq\label{generating}
	e^{F(\lambda,t)} = \sum_{e_0,e_t} \mathbb{P}(e_0,e_t) e^{\lambda E(t)}
	= \Tr\lt(e^{\ii Ht} e^{\lambda \Delta H} e^{-\ii H t} e^{-\lambda \Delta H}\rho_0\rt) =
	\Tr\lt(e^{\lambda \Delta H(t)} e^{-\lambda \Delta H}\rho_0\rt).
\eeq
We are interested in the scaled cumulant generating function (SCGF),
\beq\label{F}
	F(\lambda) = \lim_{t\to\infty} t^{-\alpha} F(\lambda,t),
\eeq
where the scaling exponent $\alpha$ is that which makes the result finite and generically nonzero (extensive variables, as is expected of the current and as arises in CFT, have exponent $\alpha=1$). Note that the function $F(\lambda)$, for $\lambda\in\R$, is always convex:
$a F(\lambda_1) +(1-a) F(\lambda_2) \geq F(a\lambda_1+(1-a)\lambda_2)$ for all $a\in[0,1]$ (as a consequence of H\"older's inequality, see e.g. \cite{Touch}).

The aim of the following is to derive an exact formula for $F(\lambda)$ for non-equilibrium homogeneous CFTs, eqs. (\ref{Fj},\ref{FCFT}) below.

Observe that
\beqs
	\Delta H(t) = \Delta H+ \int_0^t \dd s\,j(0,s).
\eeqs
Using chiral evolution \eqref{evol}, we have
\beq\label{jT}
	j(0,s) = T(-s)-\b T(s).
\eeq
That is, the time-integrated energy current is an observable supported on the interval $[-t,t]$. Thus, by the Baker-Campbell-Hausdorff formula, at each order in $\lambda$ and for every finite $t$ the average in the generating function \eqref{generating} is of an observable that is finitely supported. This is a phenomenon that is expected to be general as a consequence of the Lieb-Robinson bound. One could then take the large-$t$ limit order by order to obtain $F(\lambda)$.

However, instead of using the BCH formula, it is more convenient to argue for the result by taking the derivative with respect to $\lambda$, and writing the result in the form
\beqs
	\frc{\p }{\p \lambda} F(\lambda,t) =
	\int_{0}^{t}
	\dd s\,\frc{\Tr\lt(e^{\lambda \Delta H(t)} j(0,s)
	e^{-(\lambda+\delta) \Delta H}e^{-\b\beta (H_l+H_r)}\rt)}{
	\Tr\lt(e^{\lambda \Delta H(t)}
	e^{-(\lambda + \delta)\Delta H} e^{-\b\beta (H_l+H_r)}\rt)},
\eeqs
where we define
\beqs
	\b\beta = \frc{\beta_r+\beta_l}2,\quad \delta = \beta_r-\beta_l.
\eeqs

We first argue that we may make the replacement $H_l+H_r\mapsto H$ in the above formula. Recall that the difference $\delta H$, as for instance in \eqref{HdH}, is a locally-supported operator, which, after scaling the Hamiltonian into its integrated form \eqref{Hint}, is proportional to the infinitesimal displacement $\dd x$: we may expect it to make a vanishingly small difference in the above average\footnote{The main difference between $H_l+H_r$ and $H$ is in the corresponding time evolution: in the former, the boundary at $x=0$ is felt and reflections occur there, while in the latter, there is no such boundary and evolution is chiral throughout.}. Therefore, shifting time by $-t/2$,
\beq\label{Fgh}
	\frc{\p }{\p \lambda} F(\lambda,t) =
	\int_{-t/2}^{t/2}
	\dd s\,\frc{\Tr\lt(e^{\lambda \Delta H(t/2)} j(0,s)
	e^{-(\lambda+\delta) \Delta H(-t/2)}e^{-\b\beta H}\rt)}{
	\Tr\lt(e^{\lambda \Delta H(t/2)}
	e^{-(\lambda + \delta)\Delta H(-t/2)} e^{-\b\beta H}\rt)}.
\eeq

Second, we argue that at large times, the main contributions to the $s$-integral are those coming from its ``bulk'', and that all these contributions are equal. This is in the spirit of the existence of the large-time, steady-state limit: whenever the time at which the local observable $j(0,s)$ is evaluated is far from the time at which the density matrices are evaluated, the result is time independent. Therefore, we find an exponent $\alpha=1$ in \eqref{F}, and
\beq
	\frc{\p }{\p \lambda} F(\lambda) = \frc{\Tr\lt(e^{\lambda \Delta H(t/2)} j(0)
	e^{-(\lambda+\delta) \Delta H(-t/2)}e^{-\b\beta H}\rt)}{
	\Tr\lt(e^{\lambda \Delta H(t/2)}
	e^{-(\lambda+ \delta)\Delta H(-t/2)} e^{-\b\beta H}\rt)}.
\eeq

Finally, we argue for the form of the density matrices involve. Note that
\beq\label{Hs}
	\Delta H(s) = \frc12 \int_{-s}^s \dd x\,p(x)
	+ \int_{-\infty}^{-s} \dd x\,h(x) + \int_{s}^{\infty} \dd x\,h(x).
\eeq
We argue that, for $|s|$ large, the last two terms may be neglected in the average on the right-hand side of \eqref{Fgh}. This is easily justified by noting that, upon expanding the exponentials, one obtains thermal connected correlation functions which exponentially vanish at large distances. Using $\lim_{s\to\infty}  \int_{-s}^s \dd x\,p(x) = P$ and $\lim_{s\to-\infty}  \int_{-s}^s \dd x\,p(x) = -P$, we obtain
\beq\label{Fj}
	\frc{\p }{\p \lambda} F(\lambda) = \frc{\Tr\lt(j(0)
	e^{-\b\beta H+(\lambda+\delta/2)P}\rt)}{
	\Tr\lt(
	e^{-\b\beta H+(\lambda+\delta/2)P}\rt)} = \bra j\ket_{\mathrm{sta}}\big|_{
	\beta_l-\lambda;\beta_r+\lambda}.
\eeq
In the last equality, we re-expressed the result as an average of the energy current in the steady state with inverse temperatures shifted by the generating parameter $\lambda$.

Although the arguments above do not form a full derivation, each step was in fact proved in \cite{BD-long,BDM2015} using explicit expansions in $\lambda$ based on a Virasoro diagramatics developed there. That is, the above gives the full Taylor expansion of $F(\lambda)$.

Integrating \eqref{Fj} and using the steady-state current \eqref{jsta}, we find the energy SCGF
\beq\label{FCFT}
	F(\lambda) = \frc{c\pi}{12} \lt\{\begin{aligned}
	& \frc{\lambda}{\beta_l(\beta_l-\lambda)} - \frc{\lambda}{\beta_r(\beta_r+\lambda)} & (\lambda\in[-\beta_r,\beta_l]) \\
	& \infty & \mbox{(otherwise).}
	\end{aligned}\rt.
\eeq
Here, in order to fix $F(\lambda)$ beyond the interval $[-\beta_r,\beta_l]$, we have taken into account that poles occur at $\lambda=-\beta_r$ and $\lambda=\beta_l$, where $F(\lambda)$ is infinite, along with the convexity of $F(\lambda)$.

\subsection{Interpretations and fluctuation relations}

Re-tracing some of the steps, we note that the relation \eqref{Fj}, hence the SCGF \eqref{FCFT}, can also be obtained from the simpler definition
\beq
	F(\lambda) = \lim_{t\to\infty} t^{-1} \log\Tr\lt(\exp\lt[
	\lambda\int_0^t \dd s\, j(0,s)\rt]\,\rho_{\mathrm{sta}}\rt).
\eeq
This definition in fact gives the scaled cumulants simply as connected correlation functions of time-integrated local currents in the stationary state: $F(\lambda) = \sum_{n=0}^\infty \frc{1}{n!}\, c_n\, \lambda^n$, with
\beq\label{cumul}
	c_n =
	\lim_{\ep\to 0^+}
	\int_{-\infty}^\infty \dd s_1\cdots \dd s_{n-1}\,
	\bra j(0,s_{n-1}+(n-1) \ii\ep)\cdots j(0,s_{1} + \ii\ep)j(0,0)\ket_{\mathrm{sta}},
\eeq
where an imaginary time-ordering has been introduced in order to avoid singularities (this implements operator ordering). Tedious calculations using the steady-state correlation functions \eqref{corrfctsta} and performing the integrals in the above expression of $c_n$ show that the Taylor expansion of the function \eqref{FCFT} is indeed recovered. This explains formula \eqref{fluctuHeis} for the fluctuation cumulants in the Heisenberg chain.

We make the following further observations about the large-deviation theory underlying the above results.

\noindent (i) {\em Universality.} The evaluation of $F(\lambda, t)$ poses many more problems than that of the limit $F(\lambda)$: as observed in \cite{BD-long,BDM2015}, it contains ultraviolet singularities, pointing to the non-universality of $F(\lambda,t)$ at finite $t$. The SCGF $F(\lambda)$ is, however, fully universal in the sense of the renormalization group, and only depends on the central charge of the CFT.

\noindent (ii) {\em Large-deviation function.} Expression \eqref{FCFT} is everywhere differentiable on $\R$. Hence the time-averaged energy transfer  $J_t:=t^{-1}E(t)$ satisfies a large-deviation principle. This  says that the probability density $\mathbf{p}(J_t=J)$ that the random variable $J_t$ takes a certain value $J$ peaks, an decays exponentially in time $t$ away from the peak,
\beq
	\mathbf{p}(J_t=J)\approx e^{-tI(J)}.
\eeq
The function $I(J)$ is the large-deviation function, and can be evaluated as the Legendre-Fenchel transform of $F(\lambda)$,
\beq
	I(J) = \sup_{\lambda\in\R} \{\lambda J - F(\lambda)\}.
\eeq
As a consequence, $I(J)$ is strictly convex. Clearly, $I(J)$ is non-negative and has a zero at $J = \bra j\ket_\mathrm{sta}$, which indicates that the probability function for the random variable $t^{-1}E(t)$ concentrates, as $t\to\infty$, on the average steady-state current. This value is the unique minimum of $I(J)$, showing that the steady state, as a thermodynamic state, is unique. We see that $-I(J)$ plays the role of an entropy in this non-equilibrium setup, being maximized at the unique steady-state value (this is one way in which large-deviation theory allows for a generalization of equilibrium thermodynamics).

\noindent (iii) {\em Poisson processes interpretation.} We remark that expression \eqref{FCFT} can be written as
\beq
	F(\lambda) = \frc{c\pi}{12} \lt(
	\int_0^{\infty} \dd q\, e^{-\beta_l q} (e^{\lambda q}-1)
	+\int_{-\infty}^{0} \dd q\, e^{\beta_r q} (e^{\lambda q}-1)\rt).
\eeq
Since $e^{\lambda q}-1$ is the large-time SCGF for a single Poisson process of increments $q$, this means that the conformal SCGF is that of a combination of independent Poisson processes. It describes independent transfers of energy packets $q\in\R$, weighted in proportion to the Boltzmann weight associated to the left thermal bath (for $q>0$) or the right thermal bath (for $q<0$). This gives a natural physical interpretation to the formula \eqref{FCFT}.

\noindent (iv) {\em Extended fluctuation relations.} Recall formula \eqref{Fj}, whose integral form is
\beq\label{EFR}
	F(\lambda) = \int_0^\lambda \dd z\,
	\bra j\ket_{\mathrm{sta}}\big|_{\beta_l-z;\beta_r+z}.
\eeq
This relates the SCGF $F(\lambda)$ to the average current $\bra j\ket_{\mathrm{sta}}\big|_{\beta_l-z;\beta_r+z}$ associated with reservoirs at shifted inverse temperatures $\beta_l-z$ and $\beta_r+z$. It is equivalent to a recursion relation between scaled cumulants:
\beq
	F(\lambda) = \sum_{n=0}^\infty \frc{c_n}{n!} \lambda^n,
	\quad \Big(\frc{\p}{\p\beta_r} - \frc{\p}{\p\beta_l}\Big) c_n= c_{n+1}.
\eeq

In \cite{BD-fluctu-2013} formula \eqref{Fj} was referred to as extended fluctuation relations (EFR), and shown to hold in much more generality. It was shown to hold whenever there is ``pure transmission'', which formally is the scattering condition
\beq\label{PT}
	S^\star(\Delta H) = - S(\Delta H),
\eeq
readily verified in the CFT derivation above as a consequence of the argument around \eqref{Hs}. A different proof, not explicitly based on this scattering condition but essentially equivalent, was also given in \cite{Nat-Phys} under the assumption of PT (parity-time) symmetry of the steady-state density matrix, and of the vanishing of certain large-distance and large-time correlations. As recalled in subsection \ref{sectbeyondint}, beyond the context of CFT, the EFR was explicitly verified to hold, for the present non-equilibrium setup, in non-equilibrium free-particle models, and is conjectured to be valid for energy transport (and transport of any other higher-spin local conservation charge acting trivially on internal degrees of freedom) in every integrable model.

The EFR has many interesting consequences. From \eqref{EFR}, we see that the average non-equilibrium current gives rise to the full non-equilibrium fluctuation spectrum. But also, the full equilibrium fluctuation spectrum gives rise to the average non-equilibrium current:
\beq
	\bra j\ket_{\mathrm{sta}}\big|_{\beta-\lambda;\beta+\lambda}
	= \frc{\p}{\p\lambda}F(\lambda)\Big|_{\beta;\beta}
\eeq
where, in the indices, the first parameter represents the inverse left temperature, and the second, the inverse right temperature of the asymptotic reservoirs. At leading order in $\lambda$, this is a relation between the linear conductivity and the equilibrium noise. The linear conductivity can naturally be defined as the derivative of the current $\bra j\ket_{\mathrm{sta}}$ with respect to a change of its ``potential'' $(\beta_r-\beta_l)/2$ (keeping $\beta_l+\beta_r$ fixed), $G(\beta) =  \big(\p\bra j\ket_{\mathrm{sta}}/\p\lambda\big|_{\beta-\lambda;\beta+\lambda}\big)_{\lambda=0}$, while the noise is simply the second cumulant of energy transfer, $N(\beta) = \big(\p^2F(\lambda)/\p\lambda^2\big|_{\beta;\beta}\big)_{\lambda=0}$.
A consequence of the EFR is $N(\beta) = G(\beta).$

Further, the EFR \eqref{EFR}, combined with anti-symmetry of the current $\bra j\ket_{\mathrm{sta}}\big|_{\beta_l;\beta_r} =  -\bra j\ket_{\mathrm{sta}}\big|_{\beta_r;\beta_l}$, imply fluctuations relations of the Cohen-Gallavotti type \cite{fluctu,Jarz,Espo}:
\beq\label{FR}
	F(\beta_l-\beta_r-\lambda) = F(\lambda).
\eeq
These fluctuation relations indicate that the probability density for transfer of energy to occur in the ``wrong'' direction is exponentially suppressed as compared to that that is occurs in the ``right'' direction:
\[
	\frc{\mathbf{p}(J_t=J)}{\mathbf{p}(J_t=-J)}
	\approx e^{t(\beta_r-\beta_l)J}.
\]
Similarly, the stronger EFR is related to the following, stronger statement of exponential decay of probabilities, which can be expressed using the large-deviation function:
\[
	I(J) = \frc{\beta_l-\beta_r}2J + I_{\rm even}(\beta_l+\beta_r,J).
\]
As compared to the usual fluctuation relations, this has the additional information that the even part of the large-deviation function $I_{\rm even}(J):=(I(J)+I(-J))/2$ depends on the temperatures only as a function of $\beta_l+\beta_r$.

Finally, the question as to if there is a Poisson-process interpretation for a given $F(\lambda)$ satisfying the EFR is rephrased into the question of the positivity of the Poisson density $\omega(q)$, expressed as a Fourier transform of the stationary current:
\beq\label{Poisson}
	F(\lambda) = \int_{-\infty}^\infty \dd q\,\omega(q)\,(e^{\lambda q}-1),\quad
	\omega(q) = \frc1{q} \int_{-\infty}^\infty \frc{\dd \xi}{2\pi} e^{-\ii\xi q}\,
	\bra j\ket_{\mathrm{sta}}\big|_{\beta_l-\ii\xi;\beta_r+\ii\xi}.
\eeq
Positivity of $\omega(q)$ was verified for instance, besides CFT, for energy transport in free-particle models \cite{doyonKG}, and from conjectured non-equilibrium steady states in integrable models \cite{castroint}.

\noindent (v) {\em A greater universality.} The form \eqref{FCFT} of the SCGF was derived above using CFT. However, it is in fact a more general SCGF, which can be seen as a consequence of basic transport principles combined with the requirement that the fluctuation relation \eqref{FR} holds. Indeed, assume the following:
\bi
\item[(1)] The SCGF $F(\lambda)$ is a sum of two independent, equivalent contributions: $f(\lambda,\beta_l)$ and $f(-\lambda,\beta_r)$. These represent independent fluctuations of energy transfer coming from the left and right reservoirs.
\item[(2)] The current is proportional to the square of the temperature, $f(\lambda,\beta) \propto \lambda \beta^{-2} + O(\lambda^3)$.
\item[(3)] The FR \eqref{FR} holds.
\ei
Then \cite{BD2012} the unique solution to these requirements, up to a normalization, is \eqref{FCFT}. That is, this form of $F(\lambda)$ requires very little input from the dynamics of the transport processes, if the fluctuation relations \eqref{FR} are assumed to hold. Of course, the full CFT derivation provides a proof of these fluctuation relations. But the above remark suggests the possibility that \eqref{FCFT} may be applicable in more general situations.

\newcommand{\jq}{\mathfrak{j}^q}
\newcommand{\fq}{\mathfrak{f}^q}

\section{Charge transport} \label{sectcharge}

We may of course extend the previous discussions to the study of charge transport if the low-energy CFT possesses a conserved $u(1)$ charge current \cite{BD-long}. Charge transport was the object of some of the earliest studies of non-equilibrium steady states. It has recently received a large amount of attention, especially in the context of one-dimensional Luttinger like systems \cite{Mintchev,irrel-out,Gutman}, or of transport through Kondo or related impurities \cite{DA2006,RLM,Ben07}. Such ``impurity models'' have effective one-dimensional descriptions where baths are free fermions and the impurity produces nontrivial reflections and transmission (generically not quadratic in the free fermion fields); there, the partitioning approach is naturally implemented within the Keldysh formalism. Charge transport statistics for free-fermion models with free-fermion impurities, where both reflection and transmission occurs, is described in general by the Levitov-Lesovik formula \cite{Lev1}, see also \cite{Lev-Leso-extra}. This formula was obtained by assuming an indirect-measurement protocol, but was shown to emerge also from the two-time measurement protocol in the resonant-level model \cite{BD-rlm}. In certain non-trivial impurity models, for instance representing point-contact interactions between quantum Hall edge modes, the full statistics of charge transfer was obtained using integrability methods \cite{FLS95}. Charge transport has also been studied in star-graph configurations, where many reservoirs interact at a point \cite{Mintchev}. In homogeneous critical models for interacting fermions, it was studied in the context of Luttinger liquids (which is a CFT models with central charge $c=1$) in \cite{Gutman}. Recently, charge transport and its statistics in the homogeneous free massive Dirac theory in three dimensions was studied \cite{YDprepa}, where the Lesovik-Levitov formula was shown again to emerge from two-time measurement protocols.

For the present discussion, we concentrate on homogeneous systems (we will consider the presence of impurities, or defects, in section \ref{sectdefects}), and we may consider the same protocol as above. In order to produce a non zero charge flow, we prepare two halves of a system in Gibbs states not only at different temperatures $T_l$ and $T_r$, but also at different chemical potentials $\mu_l$ and $\mu_r$. The initial state is of the form
\[ \rho_0=\rho_l\otimes\rho_r,\quad \rho_l= \mathfrak{n}\big(\, e^{-\beta_{l,r}( H_0^{\l,r}-\mu_{l,r} Q_0^{l,r})} \,\big),\]
where $H_0^{\l,r}$ are the uncoupled left/right hamiltonians and $Q_0^{l,r}$ the left/right initial charges.

In order to fix the notation, let $\mathfrak{j}^q$ be the $u(1)$ current with local conservation law $\partial_t\jq+\partial_x \fq=0$.  Chirality of the CFT imposes that $\fq$ is also conserved so that both light-cone components are chiral \cite{CFTbook}: $(\partial_t\pm \partial_x)\jq_\pm=0$ with $\jq_\pm:=\jq\pm \fq$. 
In order to avoid confusion, we adjust the notation from previous sections by adding an index `$e$' to the energy current, which we now denote by $j^e$ (we use this notation only in this section). Since this discussion of charge transport is very parallel to the previous one for energy transport, we shall be brief. See \cite{BD-long} for more details.

\subsection{Charge transport statistics}

Let us first look at charge transport, assuming that the left/right parts of the system are initially at different chemical potentials but identical temperature. 

The initial left/ right $u(1)$ charges are $Q_0^l=\int_{-\infty}^0 \dd x\, \jq(x)$ and $Q_0^r=\int^{+\infty}_0 \dd x\, \jq(x)$ respectively. Once the two halves have been coupled, the charge current is continuous across the contact point and charges flow through this junction. Let $Q(t)$ be the variation of the total charge after a time $t$. In a way similar to energy transport and assuming that there is no reflection at the junction (only transmission occurs), the ballistic nature of the transport implies that the charge variation is due to right movers crossing the junction from the left to the right and,  reciprocally, to left movers crossing the junction from the right to the left. Hence,
\[ Q(t) = \int_{-t}^0 \dd x\, \jq_+(x) - \int^{+t}_0 \dd x\, \jq_-(x),\]
which is the difference of the left/right variations of the charges $Q_l(t):=\int_{-t}^0 \dd x\, \jq_+(x)$ and $Q_r(t)=\int^{+t}_0 \dd x\, \jq_-(x)$. 

In a CFT at temperature $T$ and chemical potential $\mu$, the $u(1)$ current expectation is proportional to the chemical potential: $\bra \jq_\pm \ket_{\beta,\mu}=\pi\, \mu$. Hence, $\bra Q(t) \ket_\mathrm{stat} = t\, \bra \jq \ket_\mathrm{stat}$ with the mean charge current $\bra \jq \ket_\mathrm{stat}$ proportional to the difference of the chemical potentials,
\begin{eqnarray}
\bra \jq \ket_\mathrm{sta} = \pi\, (\mu_l-\mu_r)
\end{eqnarray}
as expected.

As we shall explain in the following section, the fluctuations of the charge transferred $Q(t)$ is Gaussian with mean $t\, \pi (\mu_l-\mu_r)$ and covariance $t\, \pi T$, if the two halves of the system have been prepared at identical temperature $T$.

\subsection{Mixed energy and charge transport}

Let us now look at the combined statistics of energy and charge transfer. We imagine preparing initially the two halves of the systems at different temperatures $T_{l,r}$ and different chemical potentials $\mu_{l,r}$. Let $E(t)$ and $Q(t)$ be respectively the energy and charge transferred during a time duration $t$. These extensive quantities and their statistics are specified as in section \ref{sectfluctu} by a two-time measurement protocol. Hence their cumulants generating function reads
\[ e^{F(\lambda,\nu,t)}:= \mathbb{E}\big[ e^{\lambda E(t)+\nu Q(t)}\big]= \mathrm{tr}\big( e^{\lambda \Delta H(t) +\nu \Delta Q(t)}\,e^{-\lambda \Delta H - \nu \Delta Q}\, \rho_0\big),\]
with $\Delta H= (H_r-H_l)/2$ and $\Delta Q=(Q_r-Q_l)/2$. We are interested the large deviation function defined as the large time limit
\[ F(\lambda,\nu) := \lim_{t\to\infty} t^{-1}\, F(\lambda,\nu,t) .\]

Using the same arguments and manipulations than in the previous section one may prove \cite{BD-long} that the derivatives of this large deviation function is related the mean currents but shifted effective temperatures and chemical potentials depending on the formal parameters $\lambda$ and $\nu$. Namely, one has
\begin{eqnarray} \label{eq:fluctu-diff1}
\partial_\lambda F(\lambda,\nu) &=& \bra j^e \ket_\mathrm{sta}\big\vert_{\hat \beta_{l,r}(\lambda,\nu);\, \hat \mu_{l,r}(\lambda,\nu)},\\
\partial_\nu F(\lambda,\nu) &=& \bra \jq \ket_\mathrm{sta}\big\vert_{\hat \beta_{l,r}(\lambda,\nu);\, \hat \mu_{l,r}(\lambda,\nu)},
\label{eq:fluctu-diff2}
\end{eqnarray}
with
\begin{eqnarray*}
\hat \beta_{l}(\lambda,\nu)=\beta_l-\lambda &;& \hat \beta_{r}(\lambda,\nu)=\beta_r+\lambda ,\\
\hat \mu_{l}(\lambda,\nu)=\frac{\beta_l\mu_l+\nu}{\beta_l-\lambda} &;& \hat \mu_{r}(\lambda,\nu)=\frac{\beta_r\mu_r-\nu}{\beta_r+\lambda}  .
\end{eqnarray*}
These two differential equations are compatible by construction. They can be integrated to give the explicit exact form of the large deviation function, that is: 
\begin{eqnarray}
F(\lambda,\nu) = f(\lambda,\nu\vert\beta_l,\mu_l) + f(-\lambda,-\nu\vert\beta_r,\mu_r) \label{SCGFcharge}
\end{eqnarray}
with
\begin{eqnarray}
f(\lambda,\nu\vert\beta,\mu) = \frac{c\pi}{12} \lt(\frac{1}{\beta-\lambda} -\frac{1}{\beta} \rt) 
+ \frac{\pi}{2} \lt( \frac{(\beta\mu+\nu)^2}{\beta-\lambda} - \beta\mu^2 \rt).
\end{eqnarray}
This codes for both energy and charge cumulants. It is clear from this formula that charge and energy transferred are not statistically independent. Again, one must take the convex envelope of the above result, thus $F(\lambda,\nu)=\infty$ for $\lambda\in\R\setminus (-\beta_r,\beta_l)$.

A few comments are in order:
\begin{itemize}
\item The remarkable relations (\ref{eq:fluctu-diff1},\ref{eq:fluctu-diff2}) imply fluctuation relations of the Cohen-Gallavotti type,
\[ F(\lambda,\nu)=F(\beta_l-\beta_r-\lambda, \beta_r\mu_r-\beta_l\mu_r-\nu),\]
but they are stronger than the latter. In particular they only apply in the purely transmitting cases that we have discussed here. Relations (\ref{eq:fluctu-diff1},\ref{eq:fluctu-diff2}) are the extended fluctuation relations in the case of combined charge and energy transport.

\item The extended fluctuation relations (\ref{eq:fluctu-diff1},\ref{eq:fluctu-diff2}) imply the compatibility conditions
\beq
	\lt(\frc1{\beta_l}\frc{\p}{\p\mu_l}
	-\frc1{\beta_r}\frc{\p}{\p\mu_r}\rt) \bra j^e\ket_\mathrm{sta}
	=
	\lt(-\frc{\p}{\p\beta_l}+\frc{\p}{\p\beta_r}
	+\frc{\mu_l}{\beta_l}\frc{\p}{\p\mu_l}
	-\frc{\mu_r}{\beta_r}\frc{\p}{\p\mu_r}
	\rt) \bra \jq\ket_{\rm sta}.
\eeq
These may be understood in the light of Onsager's reciprocal relations \cite{onsager}: the derivative operator on the left-hand side may be interpreted as a variation of the force generating the charge current, while that on the right-hand side as the variation of the force generating the energy current.
\item The large deviation function factorizes as the sum of two contributions, one from the left half plus one from the right half of the system respectively.
\item Restricting to the energy cumulants by setting $\nu=0$, and in the case $\beta_l\mu_l = \beta_r\mu_r=:\chi$, we observe that the large deviation function is exactly that for energy transport \eqref{FCFT}, with a shifted central charge $c\mapsto c^\star = c+6\chi^2$. That is, the presence of chemical potentials and temperatures whose combined influence is exactly balanced on the right and left, has the effect of increasing the intensity of the underlying Poisson process for energy transport.
\item Restricting to the charge cumulants by setting $\lambda=0$, we recover the fact that the charge fluctuations are Gaussian with mean $\pi(\mu_l-\mu_r)$ and covariance $\pi(T_l+T_r)/2$. Indeed
\[ F(\lambda=0,\nu)=  \frac{\pi}{2}\frac{(\beta_l\mu_l+\nu)^2- (\beta_l\mu_l)^2}{\beta_l} 
+ \frac{\pi}{2}\frac{(\beta_r\mu_r-\nu)^2- (\beta_r\mu_r)^2}{\beta_r}. \]
It was shown \cite{BD-long} that this SCGF can be obtained in the two-time measurement protocol both with $t_0=0$ (as per the derivation in subsection \ref{ssectSCGF}), and with $t_0\to\infty$ (that is, with the first measurement time within the steady state itself).
\item A chemical potential difference induces a nonzero mean energy flow, even in absence of temperature difference,
\[ \bra j^e \ket_\mathrm{sta} = \frac{c\pi}{12}(T_l^2-T_r^2) + \frac{\pi}{2}(\mu^2_l-\mu_r^2) \]
as expected. But there is no thermoelectric effect in the sense that there is no non-zero charge current induced by a temperature difference in the absence of a chemical potential difference. This is because both positive and negative charge carriers have the same thermal distribution.
\end{itemize}

\section{The effects of defects} \label{sectdefects}

In all the previous examples the flows of energy or charge were totally transmitted through the contact region. There was no reflection because the total systems were homogeneous. Reflections and partial transmissions will be present if instead we introduce defects at the contact point. As already mentioned, for CFT represented by free fermions, but with nontrivial impurities, the partitioning approach for non-equilibrium steady states (usually referred to as the Keldysh formalism within this context) has been studied in a very large body of works, see e.g. the book \cite{fermion-Keldish} and references therein; generically, in impurity models, the impurity itself breaks conformal invariance. The first study of the partitioning approach within nontrivial conformal defect models is found in \cite{BDV2014}, later extended in \cite{Tauber}. The aim of this section is to briefly describe the former work: how to describe non equilibrium CFTs with partially reflecting and transmitting defects. In order to preserve the conformal structure of the overall system, the defects possess specific properties, which mainly impose compatibility with energy conservation. Once these properties have been identified and formulated, the construction of non-equilibrium CFT with defects parallels that without defects.

\begin{figure}[t]
\centering
\begin{tikzpicture}[scale=1.0]
\draw[thick,->](-1,-1)--(-0.5,-0.5);
\draw[thick](-0.5,-0.5)--(0,0);
\draw[thick,->](1,-1)--(0.5,-0.5);
\draw[thick](0.5,-0.5)--(0,0);
\draw[thick](-0.5+1,-0.5+1)--(0+1,0+1);
\draw[thick,->](-1+1,-1+1)--(-0.5+1,-0.5+1);
\draw[thick,->](1-1,-1+1)--(0.5-1,-0.5+1);
\draw[thick](0.5-1,-0.5+1)--(0-1,0+1);
\draw[yellow, fill=yellow!50] (-0.1,-1) rectangle (0.1,1);
\draw[->](-2,-1)--(2,-1);
\draw(2,-1) node[above] {$x$};
\draw[->](-2,-1)--(-2,-0.5);
\draw(-2,-0.5) node[left] {$t$};
\draw(-1,-1) node[below]{$\bar{V}^l$};
\draw(1,-1) node[below]{$V^r$};
\draw(-1,1) node[above]{$V^l$};
\draw(1,1) node[above]{$\bar{V}^r$};
\draw(-0.1,0) node[left]{$\Theta$};
\end{tikzpicture}
\caption{{\it The defect  is assumed to allow for a scattering picture of local fields belonging to the vector spaces of chiral left/right-movers. The notation is $V^{l,r}= [{}^\mathrm{left}_\mathrm{movers}]_{\mathrm{CFT}_{l,r}}$ and $\bar V^{l,r}= [{}^\mathrm{right}_\mathrm{movers}]_{\mathrm{CFT}_{l,r}}$.}}
\label{figevol}
\end{figure}
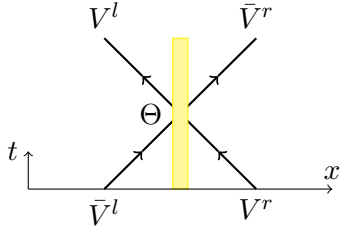

\subsection{Dynamics with defects}

In order to construct the steady state, we first need to study the dynamics in the presence of a defect, because the steady state is obtained as a long-time limit. We assume that the defect does not break scale invariance. By locality, the dynamics is chiral away from the impurity: $T(x,t)=T(x-t)$ and $\bar T(x,t)=\bar T(x+t)$ as long as the translated fields ``do not hit the defect'' (that is, both $x$ and $x-t$ have the same sign ($T(x,t)$), or both $x$ and $x+t$ have the same sign ($\b T(x,t)$)). This similarly applies to any chiral fields (such as chiral symmetry currents), $\phi(x,t)=\phi(x-t)$ and $\bar \phi(x,t)=\bar \phi(x+t)$ away from the defect, for left and right moving fields respectively. For values of $t$ such that the translated chiral field would cross the impurity, the dynamics is modified. The main hypothesis is based on a `field scattering picture': upon infinitesimal time evolution, it is possible to identify right-moving fields just on the right side of the impurity with combinations of left-moving fields on the right side of the impurity and right-moving fields on the left side;  a similar identification holds for left-moving fields just on the left side of the impurity. Since, with this prescription and upon forward chiral time evolution, left (right)-moving fields are never positioned just on the right (left) of the impurity, no other identification is necessary in order to define the evolution map. The argument is extended to the simultaneous presence of right-moving and left-moving fields on the right and left, respectively (see Figure \ref{figevol}). The scattering on the impurity is then encapsulated into a map that takes products of right-movers from the left CFT times left-movers from the right CFT into linear combination of products of left-movers from the left CFT times right-movers from the right CFT. That is:
\beqa  \label{defTheta}
\Theta: [{}^\mathrm{right}_\mathrm{movers}]_{\mathrm{CFT}_\mathrm{left}} \otimes [{}^\mathrm{left}_\mathrm{movers}]_{\mathrm{CFT}_\mathrm{right}} 
\to [{}^\mathrm{right}_\mathrm{movers}]_{\mathrm{CFT}_\mathrm{right}} \otimes [{}^\mathrm{left}_\mathrm{movers}]_{\mathrm{CFT}_\mathrm{left}}.
\eeqa
This map is called the ``defect scattering map".

If we assume that the defects do not possess any internal degrees of freedom, the CFT energy should be conserved. Recall the local conservation law $\partial_t h(x,t)+\partial_x p(x,t)=0$ with hamiltonian density $h(x,t)=T(x,t)+\bar T(x,t)$ and momentum $p(x,t)=T(x,t)-\bar T(x,t)$.
Conservation of the energy requires that the momentum operator $p(x,t)$ is continuous at the origin. This condition, which is equivalent to conformal invariance, reads
\begin{equation} \label{conserv}
 T(0^-,t)+\bar T(0^+,t)= T(0^+,t)+\bar T(0^-,t).
 \end{equation}
Pure reflection corresponds to $T(0^\pm,t)= \bar T(0^\pm,t)$, while pure transmission amounts to demanding the continuity of $T$ and $\bar T$, i.e. $T(0^-,t)= T(0^+,t)$ and $\bar T(0^+,t)= \bar T(0^-,t)$.

The energy conservation law (\ref{conserv}) yields constraints on the defect scattering maps. It has been proved in \cite{BDV2014} that this condition demands that the map $\Theta$ intertwines the Virasoro actions on the space of chiral field. Namely (\ref{conserv}) is equivalent to:
\beq\label{cond}
	\Theta \circ ( \b L_n^l+ L_n^r) = (  L_n^l + \b L_n^r ) \circ \Theta,
\eeq
where $L_n^{l,r}$ (resp. $\bar L_n^{l,r}$) are the Virasoro generators acting the space of chiral (resp. anti-chiral) fields of the left/right CFTs.
This is analog to Cardy's conditions for conformal invariance on boundary Euclidean CFTs \cite{Cardy_boundary}, although here real time CFT is considered. Notice that demanding energy conservation for all times suffices to ensures conformal invariance upon assuming that the defect has no internal degrees of freedom.

The complete time evolution is fully specified by the scattering defect map and conformality. It is defined by combining the chiral evolution and the defect map. For clarity, let us repeat the argument. Away from the defect the behavior is chiral, as long as the defect has not been crossed, so that
\begin{eqnarray}\label{chiral_left}
\phi^l(x,t) &=&\phi^l(x-t),\hskip 1.0 truecm \mathrm{for}\ x<0,\ \mathrm{all}\ t,\\
\bar \phi^l(x,t)&=&\bar \phi^l(x+t),\hskip 1.0 truecm \mathrm{for}\ x<0,\ x+t<0,\nonumber
\end{eqnarray}
for $\phi^{l}$ (resp. $\bar \phi^{l}$) any chiral (resp. anti-chiral) fields of the left CFT, and
\begin{eqnarray}\label{chiral_right}
\phi^r(x,t) &=& \phi(x-t),\hskip 1.0 truecm \mathrm{for}\ x>0,\  x-t>0, \nonumber\\
\bar \phi^r(x,t)&=&\bar \phi(x+t),\hskip 1.0 truecm \mathrm{for}\ x>0,\  \mathrm{all}\ t. 
\end{eqnarray}
for $\phi^{r}$ (resp. $\bar \phi^{r}$) any chiral (resp. anti-chiral) fields of the right CFT. 
When the fields cross the defect, their time evolution is dressed by the defect scattering map, so that
\begin{equation} \label{ThetaAB}
\bar \phi^l_a(-x,t)\, \phi^r_b(x,t) = \sum_{cd} \Theta_{ab}^{cd}\ \phi^l_c(x-t)\, \bar \phi^r_d(t-x),\quad  \mathrm{for}\  t>x>0.
\end{equation}
Here we include, in both families $\{\phi_a^{l,r}\}$ and $\{\b \phi_a^{l,r}\}$, the identity field ${\bf 1}$ itself.  We shall give simple examples in the following section. A algebraic construction of defect scattering maps was given in \cite{BDV2014}.

\subsection{Non equilibrium CFTs with defects}

The protocol described in section \ref{sectgen} can be used to produce non-equilibrium steady states for coupled CFTs with defects. We start with two uncoupled CFTs on half lines prepared at different temperatures. The two CFTs can be different, with different central charges and field contents (but they certainly both contain the chiral stress-energy tensors components). The dynamics for these uncoupled CFTs is specified by a scattering map $\Theta_0$, which simply codes for the independent boundary conditions (on the left and on the right) at $x=0$. Let $U^0_t$ be the corresponding evolution operator and $\rho_0$ be the initial density matrix. These CFTs are then put into contact in a conformally invariant way. The dynamics of the coupled system is specified by another defect scattering map $\Theta$. Let $U_t$ be the corresponding evolution operator.

The non-equilibrium steady state is obtained as before by looking at the long-time limit of the time-evolved density matrix. Hence, 
provided that the limit exists, one defines $\bra \mathcal{O} \ket_\mathrm{sta}=\lim_{t\to\infty}\Tr(\rho_0\,{U_t^0}^{-1}U_t[\mathcal{O}])$, 
for any product of operators $\mathcal{O}$. By construction, if it exists, $\bra\cdots\ket_{\rm sta}$ is $U_t$-stationary. Taking the long-time limit and defining the $S$-matrix by
\beq
	S[\mathcal{O}]:= \lim_{t\to\infty} {U_t^0}^{-1}\, U_t\, [\mathcal{O}]
\eeq
(assuming the limit exists), allows us to express the steady state as $\bra \mathcal{O} \ket_\mathrm{sta}= \Tr(\rho_0\, S[\mathcal{O}] )$. By duality the steady density matrix reads,
\begin{eqnarray}
 \rho_\mathrm{sta}= S^*(\rho_0).
 \end{eqnarray}
The $S$-matrix, and hence the steady state, is completely, and explicitly, defined by the two dual defect maps $\Theta_0$ and $\Theta$ respectively associated to the dynamical evolution $U_t^0$ and $U_t$. On chiral and anti-chiral fields, the $S$-matrix acts as follows
\begin{eqnarray*} 
S[ \phi(x) ] &=& \lt\{\ba{ll}
\phi(x),& \hskip 0.8 truecm(x<0) \\
({\cal E}_{-x}\circ  \Theta_0^{-1}\Theta)[ \mathbb{I}\otimes \phi ], & \hskip 0.8 truecm (x>0)
\ea\rt. \\
S[ \bar\phi(x) ] &=& \lt\{\ba{ll}
({\cal E}_{x}\circ  \Theta_0^{-1}\Theta)[ \bar\phi\otimes \mathbb{I} ],& \hskip 1.0 truecm (x<0)\\
\bar\phi(x), & \hskip 1.0 truecm (x>0)
\ea\rt. \label{Sbar}
\end{eqnarray*}
where the evaluation map $\mathcal{E}_x$ is defined by ${\cal E}_x[\phi^{l,r}] := \phi^{l,r}(-x)$ and ${\cal E}_x[\bar \phi^{l,r}] := \bar\phi^{l,r}(x)$
for $\phi^{l,r}$ (resp. $\bar \phi^{l,r}$) any chiral (resp. anti-chiral) fields.

These steady states $\rho_\mathrm{sta}$ are specified, at least formally, once the defect scattering maps $\Theta_0$ and $\Theta$ are defined. These maps describe the conformal boundary conditions of the decoupled CFTs, and how they are conformally coupled after the connection time. The constraint on these couplings are the intertwining relations (\ref{cond}). A large class of solutions of these constraints was presented in \cite{BDV2014}. This construction consists in considering two CFTs, the left and the right, which are not individually invariant under a group $G$ but whose product is. For instance, one may choose the left CFT to be a Wess-Zumino-Witten (WZW) model \cite{CFTbook} on a group $H\subset G$ and the right CFT to be the parafermionic coset theory $G/H$, and adjust the levels of each of those CFTs in order that the product CFT be isomorphic to the WZW model on $G$, and hence is $G$-invariant. Each element of $G$ induces a consistent defect scattering map.

In all cases studied so far the steady state carries a non-zero mean energy current, as expected, in the form
\[ \bra j \ket_\mathrm{sta} = \frac{\pi\, \mathcal{T}}{12}\, (T^2_l-T^2_r),\]
where  $\mathcal{T}$ denote a transmission coefficient depending on the CFT data and on the defect scattering map. That is, the current is still in the form \eqref{jlr}, and thus \eqref{Moorerel} holds.

The simplest case corresponds to the coupling of two free Majorana fermionic CFTs, each with central charge $c=1/2$. The coupled theory is a $c=1$ CFT with a $U(1)$ symmetry. The defect scattering map amounts to ``rotating'' the boundary fermions into each other according to this $U(1)$ symmetry. The equations of motion in the coupled system read
\begin{eqnarray} 
\label{dyn_free1}
 & ~ & \lt\{\ba{ll} \psi^r(x,t) := (\cos\alpha)\, \psi^l(x-t) + (\sin\alpha)\, \bar \psi^r(-x+t),\quad & (x>0,\ t>|x|),\\
\bar \psi^l(x,t) := (\cos\alpha)\, \bar \psi^r(x+t) - (\sin\alpha)\, \psi^l(-x-t),\quad &(x<0,\ t>|x|).
\ea\rt. 
\end{eqnarray}
where the parameter $\alpha$ codes for the scattering amplitude on the defect. The stress-tensor components of the fermions are $T^{l,r}= -\frac{i}{2} (\partial \psi^{l,r})\psi^{l,r}$ and $\bar T^{l,r}= \frac{i}{2} (\partial \psi^{l,r})\psi^{l,r}$. The mean energy current is $\bra j \ket_\mathrm{sta} = \Tr(\rho_0\, S(T(x)-\bar T(x)))$. A simple computation gives \cite{BDV2014},
\begin{equation*}
 \bra j \ket_\mathrm{sta} =\frac{\pi\,\cos^2\alpha}{24}\, \left(T_l^{2}-T_r^{2}\right),
\end{equation*}
with transmission coefficient $\mathcal{T}=\frac{1}{2}\cos^2\alpha$ as expected.

Going beyond the mean energy current and computing the higher moments of the energy transfer statistics remains an open problem (except in the free fermion case).

\section{Beyond one-dimensional CFT}\label{sectbeyond}

Beyond one-dimensional CFT, many relations and principles are expected to remain valid. However, the generalization of the results derived in the previous sections to systems beyond one-dimensional critical points necessitate a variety of new techniques, as in these cases chiral factorization fails. Two radically different situations may occur: when an integrable structure is available, the large amount of symmetries plays an important role in the non-equilibrium steady state, as conservation laws protect an infinite number of degrees of freedom. On the other hand, if the model is not integrable but, in a sense, ergodicity holds, with only few conservation laws, then only few degrees of freedom remain in the description of the steady state. The former situation is the most complicated, because of the need to describe the dynamics of a large number of degrees of freedom, even though in principle exact results might be obtainable. Up to now, the most well-established results are for free-particle models. The latter situation, that of non-integrable models, is more interesting, and developments have been made using the powerful ideas of hydrodynamics. Contrasting both situations is particularly revealing, as it underlines the strong effects of conservation laws in non-equilibrium physics.

\subsection{Free-particle models}\label{sectbeyondint}

In free-particle models, Fourier modes evolve independently, without scattering effects. Hence, each such mode may be seen as independently carrying energy or charge. As a consequence, a simple picture emerges, which generalizes the chiral factorization picture of CFT. That is, each mode associated to a right-moving (left-moving) velocity is thermalized with the equilibrium potentials of the asymptotic reservoir on the left (right). Indeed, a right-moving mode, for instance, may be seen as having travelled for a very large time from a position that is far inside the left half of the original disconnected system, and, since no scattering has occurred, keeps the associated equilibrium distribution. Let us denote by $\dd^d k\,n({\bf k})$ the total mode occupation observable in a cell of volume $\dd^d k$ around wave number ${\bf k}$, by $\varep({\bf k})$ the energy function of that mode (the dispersion relation), and by ${\bf v}({\bf k})=\nabla_{\bf k} \varep({\bf k})$ the associated group velocity. Then the steady-state density matrix for energy transport has the form
\beq\label{rhostafree}
	\rho_{\mathrm{sta}} = \frak{n}\Big(\exp\big[-\beta_l\int_{{\bf k}: v_1({\bf k})>0} \dd^d k\, n({\bf k}) \varep({\bf k}) -
	\beta_r\int_{{\bf k}: v_1({\bf k})<0} \dd^d k \,n({\bf k}) \varep({\bf k})\big]\Big)
\eeq
where $v_1({\bf k})$ is the component of the group velocity in the direction of the energy flow in the setup. Similar expressions will hold when chemical potentials are present. This density matrix naturally generalizes the form \eqref{rho_stalr} of the non-equilibrium CFT density matrix.

The picture described above is of course heuristic. In particular, Fourier modes are not localized, while some aspect of locality is assumed in order to separate left and right reservoirs. The picture can be made more precise, and the result \eqref{rhostafree} derived or even mathematically proven, by an in-depth analysis of the full time evolution of $\rho_0$ using free-fermion or free-boson techniques. This was done in models of free fermions hopping on a chain \cite{Tasaki}, in spin chain models with free fermionic representations \cite{steady-rigorous}, in the Klein-Gordon quantum field theory in arbitrary dimension \cite{doyonKG}, and in tight-binding fermionic chains, where the full time-evolution dynamics has been studied with some precision using various methods \cite{VSDH15}. It can also be understood within a semi-classical approach, as has been done in higher-dimensional free-fermion models \cite{collura2}. As the density matrix \eqref{rhostafree} describes averages of products of local operators in the steady-state limit, one must make sense of it in the infinite-volume limit. At equilibrium, this is unambiguous since there is a corresponding finite-volume equilibrium state, but there is in general no physical way of defining a finite-volume non-equilibrium state that would reproduce the states studied here. In the $C^\star$-algebraic approach to quantum chains, a correct interpretation of \eqref{rhostafree} is as an ensemble of scattering states \cite{Tasaki,steady-rigorous}. More simply, in free models, it turns out that the correct prescription is the most natural one: we must take the infinite-volume limit of a finite-volume density matrix that has the form \eqref{rhostafree}, where modes ${\bf k}$ take discrete values corresponding to wave-number quantization,
\beq
	\rho_{\mathrm{sta}} = \lim_{V\to\infty}
	\frak{n}
	\Big(\exp\big[-\beta_l\sum_{{\bf k} \in K_V: v_1({\bf k})>0} N({\bf k}) \varep({\bf k}) -
	\beta_r\sum_{{\bf k}\in K_V: v_1({\bf k})<0} N({\bf k}) \varep({\bf k})\big]\Big)
\eeq
where $N({\bf k})$ are the occupation observables, and $K_V$ is the discrete set of equally-spaced wave numbers in volume $V$. The infinite-volume limit is meant to be taken when evaluating averages of (products of) local observables.

Thanks to the factorization of the density matrix \eqref{rhostafree} into right-movers and left-movers, and to their independent dynamics, it is still true, as in CFT, that the steady-state current is a difference of a function of $T_l$ and a function of $T_r$, \eqref{jlr}, in any free particle model. For the energy current, for instance, the explicit function is
\beq
	\mathfrak{J}(T) = \int \dd^d k\,g_T({\bf k})\,v_1({\bf k})\varep({\bf k})
	= \frc12  \int \dd^d k\,g_T({\bf k})\,\frc{\p}{\p k_1}
	\lt(\varep({\bf k})^2\rt)
\eeq
where $g_T({\bf k})$ is the equilibrium density function at temperature $T$. As an example, in the $d$-dimensional Klein-Gordon model of mass $m$, the energy current is found to be \cite{doyonKG}
\beq
	\bra j\ket_{\rm sta} = \frc{d\,\Gamma(d/2)\zeta(d+1)}{2\,\pi^{\frc d2+1}}\,(r_l^2 - r_r^2)
\eeq
where $r_{l,r}^2 = T_{l,r}^{d+1}$ if $m=0$, and otherwise
\beq
	r_{l,r}^2 = \frc{1}{d! \zeta(d+1)} \int_0^\infty \dd p\,\frc{p^d}{e^{\beta_{l,r} \sqrt{p^2+m^2}}-1}.
\eeq

There are however important departures from observations made in the context of one-dimensional CFT. First, in free models outside of one-dimensional criticality (either non-critical, or in higher dimensions), the presence of a range of velocities for un-dissipated excitations, instead of a single velocity, imply that the time evolution is not described by sharp shock fronts. Instead, the transition regions, that are still within a light-cone by the Lieb-Robinson bound, are large regions where energy and charge averages have power-law  dependence.

Second, linear response is broken. Evaluating energy and pressure densities at equilibrium, one can verify that the linear-response expressions \eqref{jhlinear} are {\em not} recovered at order $(T_l- T_r)$. This is an indication of the phenomenon of generalized thermalization: the local thermalization assumption at the basis of \eqref{jhlinear}, using the equilibrium equations of state, fails, because in integrable models (such as the Klein-Gordon model), there are many more conserved charges that may be involved in local thermalization, modifying the equations of state. It appears, from the above result, that these conserved charges play an important role even in the limit $T_l\to T_r$.

Third, another departure concerns the correlation functions, and the related question of the locality of the non-equilibrium density matrix. The spatial decay of correlations is {\em not} in general exponential, as the density matrix is not local. In all free-particle models, there is a simple relation between mode occupation operators $n({\bf k})$ (or $N({\bf k})$) and local observables: mode occupations are bilinear in Fourier modes of local free-fermion or free-boson observables (fields). This means that the stationary density matrix is the exponential of an operator that is a bilinear of such observables. This is also the case, of course, for all local conserved charges of free-particle models. However, because of the jump in ${\bf k}$ at $v_1({\bf k})=0$, the exponent in \eqref{rhostafree} is not a local charge: it is rather a charge with a density whose support decays algebraically, suggesting that exponential decay is broken. 

For instance, in the Klein-Gordon model in one dimension, one finds \cite{doyonKG}
\beq
	\rho_{\rm sta} = \frak{n}\lt(\exp\lt[-
	\frc{\beta_l+\beta_r}2 H - \frc{\beta_l-\beta_r}2 (P + \widehat Q)
	\rt]\rt).
\eeq
In this expression, $H$ is the Hamiltonian and $P$ the momentum, and the charge $\widehat Q$ is
\beq\label{Qop}
	\widehat Q = \int \dd x\,\dd y\,:\phi(x)\pi(y): Q(x-y)
\eeq
where $\phi(x)$ and $\pi(y)$ are, respectively, the Klein-Gordon field and its conjugate momentum field. The kernel is
\beq\label{Qdef}
	Q(x) = -m^2 \frc{{\rm sgn}(x)}{\pi}\int_0^{\pi/2} \dd \theta
	\cos^2(\theta) e^{-m|x|\sin\theta}. 
\eeq
It is a simple matter to see that the large-$|x|$ behaviour of the kernel is
\beq\label{Qx}
	Q(x) \sim -\frc{m}{\pi x} \qquad (|x|\to\infty).
\eeq
This, in particular, leads to
\beq
	\bra \phi(x)\pi(0)\ket_{\rm sta}\sim \frc1{4\pi |x|}
	\frc{\sinh\frc{(\beta_l-\beta_r)m}2}{\sinh\frc{\beta_lm}2
	\sinh\frc{\beta_rm}2} \qquad (|x|\to\infty).
\eeq
This implies that (at least some) correlation functions vanish algebraically with the distance, in contradistinction with the CFT (massless) case, where vanishing of all connected correlation functions is exponential. It is remarkable that the vanishing of certain correlations is algebraic in the massive case, and exponential in the massless case. In spin chain models with free-particle descriptions, such as the Ising and XY chains, the study of decay of correlation functions is of a similar algebraic nature for spin operators that are bilinears in fermions, but, as was shown in \cite{expo-spin}, is exponential for spin operators that have nonlocal expressions in terms of fermions.

The full fluctuation spectrum in free-particle models can be obtained by using the extended fluctuation relations (EFR) \eqref{EFR}, discussed in section \ref{sectfluctu}. As mentioned, a general derivation for the EFR was provided in \cite{BD-fluctu-2013} based on the pure-transmission condition \eqref{PT} (satisfied in free-particle models as scattering is trivial and no reflections occur). The EFR for charge transport can be independently  shown, using free-fermion techniques, in the pure-transmission limit of the resonant-level model \cite{BD-rlm}, and in the 3+1-dimensional Dirac theory \cite{YDprepa} (but we are not aware of such explicit verifications in the cases of energy transport beyond one-dimensional CFT).

For instance, the result for the energy-transport SCGF from the EFR in the $d$-dimensional massless Klein-Gordon model is
\beq\label{FSG}
	F(\lambda) = \frc{\Gamma(d/2)\zeta(d+1)}{2\pi^{1+d/2}}
	\lt(
	(\beta_l-\lambda)^{-d} + (\beta_r+\lambda)^{-d}
	-\beta_l^{-d} - \beta_r^{-d}
	\rt).
\eeq
A Poisson-process interpretation \eqref{Poisson} also holds, as in CFT, with the Poisson density
\beq
	\omega(q) = \frc{\zeta(d+1)}{2^d \pi^{\frc{d+1}2}\Gamma\lt(
	\frc{d+1}2\rt)}\,q^{d-1}\,\lt\{\ba{ll}
	e^{-\beta_l q} & (q>0) \\
	e^{\beta_r q} & (q<0)
	\ea\rt. .
\eeq
Both of these formulae have generalizations to the massive case. One can explicitly check that the cumulants provided by \eqref{FSG} agree with the integral representation \eqref{cumul} of $c_2$, through an explicit integration of the correlation function obtained by free-field methods.

The SCGF for charge transport in the 3+1-dimensional massive Dirac theory, calculated in \cite{YDprepa}, is describable using a Levitov-Lesovik formula. We note that in the massless case, in the region of analyticity containing the real-$\nu$ line, the SCGF is a polynomial of degree 4:
\beq
	F(\nu) = \frc{T_l^3}{48\pi^2} \lt(
	\lt(\mu_l\beta_l + \nu\rt)^4 +
	2\pi^2\lt(\mu_l\beta_l + \nu\rt)^2
	-\mu_l^4\beta_l^4 - 2\pi^2\mu_l^2\beta_l^2
	\rt) - (l\leftrightarrow r),
\eeq
generalizing the polynomial of degree 2 found the one-dimensional case.

Beyond these explicit calculations, a natural, fundamental question is if the density matrix for non-equilibrium transport \eqref{rhostafree} can be connected to density matrices found in studies of thermalization in homogeneous integrable models (where the final stationary state does not carry any currents). In these studies, an important concept is that of generalized Gibbs ensembles \cite{GGE} (with a refinement based on pseudolocal charges \cite{Dtherm,GGEfinal,EFreview,Pr_thisreview}): integrable models do not, generically, thermalize, but rather a generalized thermalization phenomenon occurs. That is, after an infinite-time evolution with a local, homogeneous Hamiltonian from a homogeneous state, if the limit of local observables exist, the stationary density matrix, instead of being thermal, takes the form $\exp\lt[-W\rt]$ where $W$ is a pseudolocal conserved charge of the model (often written in a basis decomposition on the Hilbert space of pseudolocal conserved charges, $W=\sum_i \beta_i Q_i$). Pseudolocal charges \cite{Dtherm,Pr_thisreview,Drude_bound,pseudolocal} generalize the usual local charges of integrable models, and are defined mainly by the fact that their second cumulants scale with the volume (further, their ``action'' should exist on local observables, and should be translation invariant)\footnote{The definition of pseudolocal charges is state-dependent \cite{Dtherm}, and thus the space of pseudolocal charges depends on the state considered.}. That such a form for the stationary density matrix occurs has been convincingly established in many examples \cite{GGEfinal,GGEex}, and it has been shown that this is a general phenomenon, in the sense that under long-time existence conditions, the stationary state has to be of this form \cite{Dtherm}. Of course, as discussed above, the density-matrix formalism, although intuitive, is not well adapted to describe steady states of thermodynamic models. A mathematical definition of generalized Gibbs ensembles directly in the thermodynamic limit was given in \cite{Dtherm}, based on clustering states (whose cumulants vanish fast enough with the distance) and on defining states via flows tangent to pseudolocal charges. In all studies of (generalized) thermalization, the generalized Gibbs ensemble does not carry any current, and one may ask if the current-carrying density matrix \eqref{rhostafree} takes the same general form.

We give here a brief answer, for instance, in the example of the one-dimensional massive Klein-Gordon model, but it is expected to be generic. Taylor expanding $\pi(y)$ around $y=x$ in \eqref{Qop} reveals that at every power of $(y-x)$, the resulting integral over $y$ is divergent: $\widehat Q$ cannot be expressed as a converging series of local conserved charges. On the other hand, let us define $\widehat Q_L$ by restricting the integration regions in \eqref{Qdef} to $[-L,L]$. Let us also consider some state $\bra\cdot\ket$ where Wick's theorem holds and $\Z_2$ symmetry is not broken (hence, for instance, one-point functions of $\phi(x)$ and $\pi(x)$ vanish), and assume the two-point functions $\bra\phi(x)\phi(y)\ket$, $\bra\phi(x)\pi(y)\ket$ and $\bra\pi(x)\pi(y)\ket$ decay at least proportionally to $|x-y|^{-1}$ at large distances. The steady state $\bra\cdots\ket_{\rm sta}$ is such a free-field, algebraically clustering state. Then, thanks to the behaviour \eqref{Qx}, one can show that the quantity $\bra \widehat Q_L^2\ket- \bra \widehat Q_L\ket^2$ grows linearly with $L$. Further, one can also show that $\lim_{L\to\infty}(\bra \widehat Q_L A\ket - \bra \widehat Q_L\ket\bra A\ket)$ exists, for $A$ any product of fields $\phi(x)$ and $\pi(x)$ at different points, and that the appropriate condition of translation invariance holds. Thus $\widehat Q$ is a pseudolocal charge with respect to that state\footnote{$\widehat Q$ is not, however, a quasi-local charge: its kernel $Q(x)$ does not decay exponentially with $|x|$.}. A pseudolocal flow \cite{Dtherm} is produced by considering the family of states $\bra \cdots\ket(s),\;s\in[0,1]$ obtained by replacing $\beta_{l,r}\mapsto s\beta_{l,r}$, and by verifying that the derivative with respect to $s$ is the finite action of this pseudolocal charge. Since the algebraic clustering in $|x-y|^{-1}$ is at the (excluded) boundary of the manifold of clustering states studied in \cite{Dtherm}, this suggests that the stationary state is (the limit of) a generalized Gibbs ensemble.

Finally, we note that the generalization of the above discussion to one-dimensional interacting integrable models appears to be natural: in such models, by elastic scattering, the individual particles' energies are conserved at every collision, and thus from the viewpoint of energy transport, similar heuristic arguments can be put forward in order to obtain density matrices of the type 
\eqref{rhostafree}. However, in interacting models, the correct finite-volume regularization may be more involved. Conjectures for steady-state density matrices were given in \cite{DeLucaVitiXXZ,castroint} using Bethe ansatz techniques, based on general ideas from scattering states \cite{doyqft}; we note that these conjectures break the property \eqref{jlr}. These are conjectures, and there is unfortunately, to date, no full derivation.

\subsection{Higher-dimensional and non-integrable QFT: Emerging hydrodynamics} \label{secthydro}

Beyond 1D critical points and beyond integrable models, new techniques must be found. One of the most powerful ideas in studying the dynamics of quantum field theory is that emerging from a hydrodynamic description of local averages \cite{Landau}. Hydrodynamics allows to encode in a simple way non-equilibrium states, including states with constant flows and approaches to steady states, by concentrating only on quantities of physical relevance and without the need for an infinite number of degrees of freedom. In the quantum context, the passage from strongly interacting many-body quantum dynamics to classical hydrodynamics involve subtle effects  \cite{Qhydro}. In this subsection, we show how to extend some of the results for non-equilibrium steady states in the partitioning approach to models that are not integrable, using the ideas of emerging hydrodynamics. We refer to \cite{hydro} for tutorials on hydrodynamic equations and their solutions.

In section \ref{sectgen} we provided arguments for a non-equilibrium scenario in the linear-response regime: sound waves are emitted from the contact point at $x=0,\,t=0$, propagating in opposite directions, and separate space-time into three regions: two equilibrium reservoirs (on the left and on the right), and a steady-state region (in-between the waves) with energy current and density \eqref{jhlinear}. We saw in section \ref{sectCFT} that the results extend naturally beyond the linear-response regime in one-dimensional CFT.

The main idea behind both the linear-response and CFT calculations of the stationary quantities is that of solving current-conservation equations. Indeed, in both cases we considered the equations
\beq\label{conslaw}
	\p_t \bra h\ket+ \p_x \bra j\ket=0,\quad
	\p_t \bra p\ket + \p_x \bra k\ket=0
\eeq
for averages of densities and currents in the evolving state $\bra \cdots\ket = \Tr\lt(\frak{n}(e^{-\ii Ht} \rho_0 e^{-\ii Ht})\cdots\rt)$, along with the constraints $\bra j\ket=\bra p\ket$ and $\bra k\ket={\cal F}(\bra h\ket)$. For relativistic invariant systems, as we argued in subsection \ref{ssectcrit}, the relation $j=p$ is an operator relation, hence holds in every state. On the other hand, the relation $\bra k\ket={\cal F}(\bra h\ket)$ holds either as an approximation coming from the assumption of local thermalization (assumed to be valid near equilibrium), or exactly as a consequence of scale invariance in one dimension, $k=h$ (subsection \ref{ssectcrit}).

A natural way forward in order to study steady states away from one-dimensional criticality is to keep the assumption of local thermalization, but without the linear-regime approximation. In order to have the possibility of ballistic transport, we consider models near criticality, so that translation-invariant collective behaviours dominate the physics (see subsection \ref{ssectcrit}). Then, at large times, one would expect microscopic relaxation processes to have occurred and every local environment to have reached some highest-entropy state. Yet, for times that are large but still shorter than the relaxation time $\tau_e$ for collective behaviours (subsection \ref{ssectpheno}), conservation laws of QFT will still hold: for instance, effects of collision with lattice phonons will not have occurred in great amounts, so that QFT collective behaviours will be intact. Thus, at such large but not infinite times, we would expect to be able to describe the state as locally thermalized: entropy is locally maximized under conditions of all QFT local conservation laws, and the associated potentials are space-time dependent. Assuming that the only conservation laws (relevant to the specific geometry at hand) are space-time translation invariance, with conserved charges $H$ (the hamiltonian) and $P$ (the momentum operator in the direction of the flow), the local-thermalization assumption is
\begin{eqnarray} \label{eq:rholocal}
\rho_{\rm local} \propto \exp\Big( - \beta(x,t)\int_{\mathfrak{N}_{x,t}}  \dd x'\, h(x')+ \nu(x,t)\int_{\mathfrak{N}_{x,t}}  \dd x'\, p(x')\Big)
\end{eqnarray}
with $x,t$-dependent inverse temperature $\beta(x,t)$ and momentum potential $\nu(x,t)$. The local region $\mathfrak{N}_{x,t}$ around $(x,t)$ where local thermalization occurs is referred to as a fluid cell.
This means that, for local observables $\Or$, the hydrodynamic-approximation average $\bra\Or\ket(x,t)$ at point $x$ and time $t$ is
\beq\label{fluidcell}
	\bra \Or\ket(x,t) = \Tr\Big(\frak{n}\big(e^{-\beta(x,t)H +\nu(x,t)P}\big)\Or\Big).
\eeq

Now consider conserved densities and currents, say with the conservation equation $\p_t h+ \p_x j =0$ and $\p_t p + \p_xk=0$ (as operator relations in QFT). The integrated version of this equation can be written on any contour in space-time. Choosing contours around regions that are much larger than the fluid cells, we deduce that the conservation equation must hold as well at this larger scale. Thus we may write
\beq\label{fluideq}
	\p_t \bra h\ket(x,t)+ \p_x \bra j\ket(x,t)=0,\quad
	\p_t \bra p\ket(x,t) + \p_x \bra k\ket(x,t)=0
\eeq
where the averages are evaluated using \eqref{fluidcell}. Note that with two conservation equations, we have two potentials describing the fluid cells. Thus these are two equations for two unknown: the potentials $\beta(x,t)$ and $\nu(x,t)$. These potentials can be taken as the hydrodynamic variables, and the resulting equations are the hydrodynamic equations associated to the QFT with space-time translation conservation laws. The problem is model-dependent through the dependence of the densities and currents on the two potentials.

Other, more physical hydrodynamic variables can be taken instead, for instance the averages $ \bra h\ket(x,t)$ and $ \bra p\ket(x,t)$. Since only two parameters ($\beta$ and $\nu$) give rise to four functions (the four averages), we may write $\bra j\ket(x,t)$ and $\bra k\ket(x,t)$ as functions of $\bra h\ket(x,t)$ and $\bra p\ket(x,t)$: these are ``generalized'' equations of state. Denoting these functions, respectively, as
\beq\label{ges}
	{\tt j} = {\cal G}({\tt h},{\tt p}),\quad {\tt k}={\cal F}({\tt h},{\tt p})
\eeq
the hydrodynamic equations become
\beq
	\lt(\begin{matrix}
	\p_t {\tt h} \\
	\p_t {\tt p}
	\end{matrix}\rt)
	+
	\lt(\begin{matrix}
	\p_{\tt h} {\cal G} & \p_{\tt p}{\cal G}\\
	\p_{\tt h} {\cal F} & \p_{\tt p}{\cal F}
	\end{matrix}\rt)
	\lt(\begin{matrix}
	\p_x {\tt h}\\
	\p_x {\tt p}
	\end{matrix}\rt)
	=0.
\eeq
We see that the Jacobian matrix
\beq
	J({\tt h},{\tt p}) = \lt(\begin{matrix}
	\p_{\tt h}{\cal G} & \p_{\tt p}{\cal G} \\
	\p_{\tt h}{\cal F} & \p_{\tt p}{\cal F}
	\end{matrix}\rt)
\eeq
for the change of coordinates from densities to currents $({\tt h},\,{\tt p})\mapsto ({\tt j},\,{\tt k})$ is involved. 

These are equations for ``pure hydrodynamics'' (without viscosity -- see below). If more conservation laws are present and relevant to the geometry of the problem, then these must be added to this hydrodynamic description. For instance, in integrable models, infinitely-many such conservation laws exist, and there is no finite hydrodynamics (the question as to a hydrodynamic description of interacting integrable models is still open).

A final ingredient to the hydrodynamic picture of non-equilibrium QFT is that concerning the family of solutions to the equations \eqref{fluideq}. We note that these equations are scale invariant: they are unchanged under the trivial scaling $(x,t)\mapsto(ax,at)$. This is related to the fact that these are effective large-scale equations: the large-scale limit has been taken already. Likewise, the solution should also possess this invariance, indicating that any short-scale structure has been washed-out by focussing on large scales. The solution will thus not capture any structure that vanishes in the scaling $(x,t)\mapsto(ax,at)$, $a\to\infty$. We are then looking for self-similar solutions, of the form
\beq
	\beta(x,t) = \beta(x/t),\quad \nu(x,t) = \nu(x/t),
\eeq
and we have, with $\xi=x/t$,
\beq\label{hydro}
	\xi \p_\xi \bra h\ket(\xi) = \p_\xi \bra j\ket(\xi),\quad
	\xi \p_\xi \bra p\ket(\xi) = \p_\xi \bra k\ket(\xi).
\eeq
The hydrodynamic equations may then be written in the form
\beq\label{hydroeq}
	(J-\xi {\bf 1})
	\lt(\begin{matrix}
	\p_\xi{\tt h}\\ \p_\xi{\tt p}
	\end{matrix}\rt) = 0
\eeq
where ${\bf 1}$ is the identity matrix. The initial conditions are then imposed via asymptotic conditions in $\xi$: the left reservoir at $\xi\to-\infty$, and right reservoir at $\xi\to\infty$.

Note that the restricting to self-similar solutions does not relate to the potential underlying scale (and conformal) invariance of the QFT model. It is simply a consequence of the hydrodynamic description of the full quantum problem, emerging at larges scales.

Solutions in general will be composed of regions of constant hydrodynamic variables (such as the reservoirs themselves), separated by transition regions. There are two types of transition regions that we must consider. The first are continuous transition regions, covering an extended interval in $\xi$, where hydrodynamic variables vary continuously between two constant regions. These are usually referred to as rarefaction waves. The second type are transition regions supported on points in $\xi$, where abrupt, discontinuous changes occur. These are referred to as shocks (or, under certain conditions, are considered to be contact discontinuities, see e.g. \cite{hydro}).

For the first, we ask for solutions that are continuous and differentiable. For such a solution, \eqref{hydroeq} indicates that the curve $\xi\mapsto ({\tt h},{\tt p})$ is tangent, at every parameter $\xi$, to an eigenvector  of the Jacobian  $J({\tt h},{\tt p})$ with eigenvalue $\xi$. If the eigenvector is nonzero, given $({\tt h},{\tt p})$, one must take a real value $\xi_-\in\R$ in the spectrum ${\rm spec}(J({\tt h},{\tt p}))$ of the Jacobian matrix, and solve for the curve starting at $\xi=\xi_-$. Since, if the solution exists, the curve is then uniquely defined, it is clear that, generically, we cannot expect a differentiable solution to the full asymptotic-value problem for the non-equilibrium steady state, as we would have to require that the curve starts and ends at specified points. However, the solution may be differentiable in various regions that are continuously connected. For instance, it may be constant up to $\xi_-$, then related to a nonzero eigenvector in $[\xi_-,\xi_+]$, and then constant again. Given $\xi_+$ (and $\xi_-$) there is a unique point $({\tt h},{\tt p})$ at the end of the curve. Such a region is called a rarefaction wave, because it is bounded by finite velocities $\xi_-$ and $\xi_+$ and it extends in space linearly with time.

The second type of solutions, that of shocks, are weak solutions: at fixed values of $\xi$, there are finite jumps, where the above hydrodynamic equations break down and the values of densities and currents are ill-defined. At these points, we cannot use the generalized equations of state. Instead, at a shock speed $u$, we may simply set
\beqa
	\p_\xi\bra h\ket(\xi) = \Delta {\tt h}\,\delta(\xi-u) &,&
	\p_\xi\bra j\ket(\xi) = \Delta {\tt j}\,\delta(\xi-u), \nonumber\\
	\p_\xi\bra p\ket(\xi) = \Delta {\tt p}\,\delta(\xi-u) &,&
	\p_\xi\bra k\ket(\xi) = \Delta {\tt k}\,\delta(\xi-u), \nonumber
\eeqa
and \eqref{hydro} gives
\beq\label{RH}
	u\Delta{\tt h} = \Delta {\tt j},\quad u\Delta {\tt p} = \Delta {\tt k}.
\eeq
These are known as the Rankine-Hugoniot conditions, and are interpreted as conservation conditions across the shock. Thus, given $({\tt h},{\tt p})$ on the left of the shock, the equations of state give $({\tt j},{\tt k})$ on the left, and, fixing $u$, the above are two algebraic equations for the values on the right of the shock.

The two types of transition regions are put together by setting the asymptotic values for ${\tt h}$ and ${\tt p}$ on the left, and keeping them constant except for rarefaction waves and shocks. At the crossing of every rarefaction wave and shock, unique changes of the values of ${\tt h}$ and ${\tt p}$ are, in principle, determined, given either the choice of initial element $\xi_-$ of the Jacobian spectrum and the right-limit of the rarefaction wave $\xi_+$, or the shock speed $u$. The problem is then to find values of $(\xi_-,\xi_+)$ and $u$ such that the asymptotic reservoir on the right is reached.

Generically, this problem has an infinity of solutions. The correct physical solution (which would come out, for instance, of a numerical study of the above hydrodynamic equations) comes from considerations of the terms omitted in taking the limit $(x,t)\mapsto(ax,at)$, $a\to\infty$. One may add viscosity terms, which involve higher derivatives and which would, in principle, be derived from a precise analysis of the QFT model at hand. These viscosity terms, being higher-order in the space-time derivatives, formally disappear after infinite scaling. However, they still affect weak solutions, as pure hydrodynamic breaks down at the positions of shocks. Viscosity terms give rise to inequalities, interpreted as entropy-production inequalities, which have to be imposed on weak solutions of pure hydrodynamics. Usually, these additional conditions uniquely fix the solution to the pure hydrodynamic problem. Without the knowledge of the correct viscosity terms, one may nevertheless impose the physical requirement of entropy production in order fix the solution.

Note that a solution with two ``structures'' (two shocks, or one shock and one rarefaction wave, etc.), describing the two transitions separating three regions (the two reservoirs, and a central region), has exactly the correct number of parameters in order to be, in principle, uniquely solved given the asymptotic reservoirs' conditions. These are the solutions we are usually looking for.

Once the solution is found, the full steady-state density matrix is obtained as
\[
	\rho_{\rm sta}=\frak{n}\Big(e^{-\beta_{\rm sta}H+\nu_{\rm sta}P_1}\Big)
\]
with the potentials $\beta_{\rm sta} = \beta(\xi=0)$ and $\nu_{\rm sta} = \nu(\xi=0)$.

We remark that, as in one-dimensional CFT but in contrast to free-particle models, this is a local density matrix, hence correlation functions display exponential spatial decay. This density matrix naturally generalizes the form \eqref{rho_sta} of the non-equilibrium CFT density matrix. We also note that the linear-response relations \eqref{jlr} are naturally recoverd at small temperature differences, as local thermalization is embedded within the hydrodynamic techniques.

\subsection{Example: higher-dimensional CFT}

Let us concentrate on relativistically invariant systems, with $j=p$. This strongly constrains the form of the locally-thermalized averages. Indeed, as explicitly shown in \cite{BDirre}, this structure leads to exact relations for averages in states that are Lorentz transforms of thermal states, involving both the Hamiltonian and the momentum operator,  $\bra \cdots\ket_\theta = \Tr \lt(\frak{n}\lt(e^{-\beta_{\rm rest}(\cosh \theta\,H - \sinh \theta\,P)}\rt)\cdots\rt)$:
\beqa
	\bra h\ket_\theta &=& \cosh^2\theta \,\bra h\ket_0 + \sinh^2\theta\,\bra k\ket_0\n
	{\bra k\ket}_\theta &=& \sinh^2\theta \,\bra h\ket_0 + \cosh^2\theta\,\bra k\ket_0\n
	{\bra p\ket}_\theta &=& \sinh\theta\cosh\theta\,\big(\bra h\ket_0 + \bra k\ket_0\big)
	\label{p1}
\eeqa
and
\beq\label{p2}
	(\bra h\ket_0 + \bra k\ket_0)\dd T_{\rm rest} = T_{\rm rest}\dd \bra k\ket_0.
\eeq
The latter is particularly useful in order to determine the temperature dependence of such averages. Therefore, locally-thermalized averages can be naturally parametrized by a rest-frame inverse temperature $\beta_{\rm rest} = \beta^2-\nu^2$ and a rapidity $\theta = {\rm arctanh}\,(\nu/\beta)$. In particular, once the zero-rapidity thermal equation of state $\bra k\ket_0 = {\cal F}(\bra h\ket_0)$ is known, averages of the energy density, the energy current and the pressure can all be evaluated as functions of the rest-frame temperature and the rapidity. This is obtained using the temperature dependence of $\bra k\ket_0$ and $\bra h\ket_0$ deduced from \eqref{p2} in the form
\beq\label{exactener}
	\log T = \int^{\bra k\ket_0} \frc{\dd \ell}{\ell + {\cal F}^{-1}(\ell)}
	= \int^{\bra h\ket_0} \frc{\dd \ell\,{\cal F}'(\ell)}{\ell+{\cal F}(\ell)},
\eeq
and inserting into \eqref{p1}. Thus, the problem is fully solved (up to a normalization of the temperature) simply by solving the equations resulting from \eqref{p1} and \eqref{fluideq}. The self-similar solutions will be obtained with the procedure described above, with the generalized equations of state determined by
\beq
	{\cal G}({\tt h},{\tt p}) = {\tt p}\quad\mbox{and}\quad
	\frc{\p}{\p\theta} {\cal F}({\tt h}_\theta,{\tt p}_\theta)
	=2{\tt p}_\theta,\quad {\cal F}({\tt h},0) = {\cal F}({\tt h})
\eeq
(where ${\tt h}_\theta = \cosh^2\theta\,{\tt h}+\sinh^2\theta\,{\cal F}({\tt h})$ and ${\tt p}_\theta = \sinh\theta \cosh\theta\,({\tt h}+{\cal F}({\tt h}))$). This holds, for instance, in higher-dimensional CFT, as well as in one- or higher-dimensional CFT perturbed by non-integrable perturbations. Models will only differ in the choice of the equilibrium equations of state ${\cal F}({\tt h})$.

Consider the example of higher-dimensional CFT. The non-equilibrium problem was solved in \cite{Nat-Phys,CKY,rarefact1,rarefact2} using hydrodynamic methods. In this case, by tracelessness of the stress-energy tensor $T^\mu_{\;\nu}=0$ and isotropy of the equilibrium thermal state, one obtains
\beq
	{\cal F}({\tt h}) = d^{-1}{\tt h}\quad \mbox{(CFT)}
\eeq
where $d$ is the dimension of space. Using the relativistic structure, equations \eqref{fluideq} may be written using the stress-energy tensor ${\tt T}^{\mu\nu}$ in the form
\beq\label{relhydro}
	\p_\mu {\tt T}^{\mu\nu}=0,\quad {\tt T}^{\mu\nu} = a T^{d+1}((d+1)u^\mu u^\nu + \eta^{\mu\nu}),\quad \eta={\rm diag}(-1,1,\ldots,1)
\eeq
where $u^\mu = (\cosh\theta,\sinh\theta)^\mu$ is the fluid velocity and $a$ is a model-dependent normalization constant. In applying the procedure outlined above, one must determine the entropy-production condition selecting the correct solution. It is a simple matter to verify that $s^\mu:=T^du^\mu$ is conserved as a consequence of the pure hydrodynamic equations \eqref{relhydro}. This is proportional to the entropy current, and, it turns out, the entropy-production condition is $\p_\mu s^\mu\geq0$. At every shock, we must require that the integrated version of this inequality holds around the shock. Amongst the two-structure solutions, it turns out \cite{rarefact1,rarefact2} that the only possible one is that where a shock enters the lowest-temperature reservoir (on the right if $T_l>T_r$), while a rarefaction wave separates the highest-temperature reservoir (on the left) from the central region -- the simpler two-shock solution would break the entropy-production condition. Surprisingly however \cite{rarefact1}, the two-shock solution, presented in \cite{Nat-Phys,CKY}, is correct up to order $\big((T_l-T_r)/(T_l+T_r)\big)^3$.

Inside the rarefaction wave, at coordinate $\xi$, the local fluid velocity $v = \tanh\theta$ is simply the relativistic sum of the sound velocity $v_s=\frc1{\sqrt{d}}$ and the ray velocity $\xi$:
\beq\label{vCFT}
	v=v(\xi) = \frc{\xi+v_s}{1+\xi v_s},
\eeq
while the local rest-frame temperature $T_{\rm rest}$ is
\beq
	T_{\rm rest}(\xi) = T_l \lt(\frc{1-v(\xi)}{1+v(\xi)}\rt)^{v_s/2}.
\eeq
The rarefaction wave starts at the sound velocity $\xi_- = -v_s$. The parameters in the central region are then functions of its end-point $\xi_+$, given by $v(\xi_+)$ and $T_{\rm rest}(\xi_+)$. This end-point, and the shock velocity $u>\xi_+$ separating the central region from the right reservoir, are then two parameters that must be fixed using the two Rankine-Hugoniot conditions \eqref{RH} around the shock at velocity $u$. The solution is unique, but there is no simple analytic expression for it. Nevertheless, a simple linear-regime analysis recovers \eqref{jhlinear}.

From \eqref{vCFT} we see that if the local fluid velocity in the central region exceeds the speed of sound, $v(\xi_+)>v_s$, then the rarefaction wave covers the steady-state ray $\xi=0$ (the ray describing the large-$t$ limit with $x$ fixed). This happens when the ratio $T_l/T_r$ if larger than a certain quantity $\Gamma$ that only depends on the dimensionality $d$ \cite{rarefact1}. In this case, the steady state, inside the rarefaction wave, is at fluid velocity $v(0)=v_s$ and rest-frame temperature $T_{\rm rest}(0) = T_l\lt(\frc{1-v_s}{1+v_s}\rt)^{1/(2\sqrt{d})}$. Surprisingly, the state then only depends on the left-reservoir temperature $T_l$. In particular, the energy current takes the simple form
\[
	\bra j\ket_{\rm sta} = a\frc{(d+1)\sqrt{d}}{d-1}
	\lt(\frc{\sqrt{d}-1}{\sqrt{d}+1}\rt)^{\frc{d+1}{2\sqrt{d}}}T_l^{d+1}
	\quad (T_l/T_r>\Gamma).
\]

We remark that the steady state in higher-dimensional CFT has a very different form for integrable and non-integrable models: in the former case, we saw that it was controlled by pseudolocal conserved quantities, while in the latter, it is solely determined by the energy and momentum potentials. This is in stark difference to the one-dimensional case, where integrability does not affect the general form \eqref{rho_sta}.

\subsection{Example: irrelevant $T\b T$ perturbation to one-dimensional CFT}

Finally, we briefly describe the example of an irrelevant perturbation of one-dimensional CFT. The perturbation to the Hamiltonian chosen in \cite{BDirre} is $g\int \dd x \,T(x) \b T(x)$, which is expected to provide a contribution to the effects of band curvatures which are felt as temperature increases. This does not takes into account incoherent effects such as those of phonon collisions, hence will not destroy the steady state. It simply gives modifications, providing higher-order expansions in $T_l$ and $T_r$ of the stationary averages.

In \cite{BDirre} it was found that this perturbation {\em does not} break relativistic invariance at order $g$, and is equivalent to the following equation of state:
\beq\label{eos}
	{\tt k} = {\cal F}({\tt h}) = {\tt h} +\frc g2{\tt h}^2 + O(g^2).
\eeq
From this, the generalized equations of state are found to be
\beq\label{geos}
	{\cal F}({\tt h}, {\tt p}) = {\tt h} +\frc g2({\tt h}^2 - {\tt p}^2) + O(g^2),\quad
	{\cal G}({\tt h}, {\tt p}) = {\tt p} + O(g^2).
\eeq
The exact temperature dependence of the energy density at equilibrium with temperature $T$ is, from \eqref{exactener},
\beq\label{hth}
	\bra h\ket = \frc{c\pi}{6} T^2\lt(1-\frc{gc\pi}{8}T^2 + O(g^2)\rt),
\eeq
and this gives a sound velocity, at temperature $T$,
\beq\label{sound}
	v_s(T) = 1 + \frc{gc\pi}{12} T^2 + O(g^2).
\eeq

One may then repeat the above analysis with this new equation of state. It turns out that the perturbative solution to the full quantum problem to order $g$ is in agreement, in the steady-state region, with the hydrodynamic approximation and a two-shock solution. We find the following \cite{BDirre}:
\bi
\item The steady-state current is still a difference of a function of the left-reservoir temperature, $\mathfrak{J}(T_l)$, minus the same function of the right-reservoir temperature, $\mathfrak{J}(T_r)$, as in \eqref{jlr}, with
\beq
	\mathfrak{J}(T) = \frc{c\pi}{12} T^2\lt(1-\frc{gc\pi}{12} T^2 \rt).
\eeq
Recall that this implied \eqref{Moorerel}.
\item The shock velocities are exactly the sound velocities \eqref{sound} of the left and right reservoir, $u_l = -v_s(T_l)$ and $u_r= v_s(T_r)$ respectively. That is, it is the linear sound velocities of the reservoirs that control the speeds of the shocks describing the steady state, as in the unperturbed and linear-response scenarios, but with appropriate dependence on the left and right reservoirs' temperatures.
\item One can check that near equilibrium (i.e. with $T_l\sim T_r$), one recovers \eqref{jhlinear}.
\ei

Finally, the $T\b T$-perturbation may also be treated using random Virasoro flows \cite{BDirre}. In this picture, the interaction, via a type of decoupling of $T$ and $\b T$, gives rise to a time evolution along stochastic vector fields. The stochasticity modifies the shock velocities and the steady state current, reproducing exactly the results shown above. It can also be used to argue for the way the shocks spread due to the microcsopic effects of band curvature (represented by the interaction), and the way the steady state is approached, in the spirit of fluctuating hydrodynamics \cite{Spohn}. This gives a shock spreading of the form $(gt)^{1/3}$ and a steady state approach with correction terms of the form $t^{-1/2}$. These are, however, conjectures.

The idea of interactions giving rise to random flows, and random flows giving rise to hydrodynamics, is worth exploring further.

\subsection{Miscellaneous general relations}

A general theory of non-equilibrium quantum steady states is still largely missing. For instance, the quantum counterpart of a macroscopic fluctuation theory \cite{MFT} has not been fully developed yet, although the results reviewed above concerning the scaled cumulant generating function can be interpreted in this light within one-dimensional CFT. The most powerful general relations, which have been verified in many models, are the fluctuation relations of Cohen-Gallavotti and Jarzynski type. A recent development is that of expansion potentials \cite{VMreview,VKM2015}, which concentrates however on total, volume integrated currents instead of steady-state properties. Here we describe two types of general relations for steady states based on simple conservation laws, which may form part of a general theory.

The first concerns the question of the existence of a nonzero current in the situation where conservation laws \eqref{conslaw} exist. If the energy current is the momentum density, $\bra j \ket = \bra p\ket$, the calculation in subsection \ref{ssectlinresp} shows that, in linear response, the current is nonzero. Can we bound the current in the full non-equilibrium regime, thus showing non-equilibrium ballistic transport? It turns out that \cite{D_2014}, under natural conditions on the behaviour of the pressure $\bra k\ket$ inside the space-time transition regions separating the asymptotic baths from the steady state, a lower bound indeed exists. These conditions are natural and the lower bound can indeed by verified in all exact results obtained to date, including free-particle and  hydrodynamic results. An upper bound for the non-equilibrium current can also be obtained by similar arguments, under natural conditions on the energy density within the transition regions. These bounds generalize the linear-response relations \eqref{jhlinear}, and in the general case, with $\bra j \ket$ and $\bra p\ket$ kept independent, they are
\beq
	v_{LR}\frc{\bra h\ket_{\beta_l} - \bra h\ket_{\beta_r}}{2}
	\geq \bra j\ket_{\rm sta},\quad \bra p\ket_{\rm sta}\geq
	\frc{\bra k\ket_{\beta_l} -\bra k\ket_{\beta_r}}{2v_{LR}}
\eeq
where $v_{LR}$ is the Lieb-Robinson velocity \cite{lr}. That is, the difference of the reservoirs' energy densities give a maximum for the steady-state current, and the different of their pressures give a minimum, in accordance with physical intuition. We note that in one-dimensional CFT, both bounds are saturated.

For the second family of relations, recall that in the hydrodynamic study of non-equilibrium steady states, an important concept is that of the generalized equations of state \eqref{ges}. This comes from density operators that include both parity-even and parity-odd conserved charges,
\beq
	\bra \Or\ket = \Tr\lt(\frak{n}\lt(e^{-\beta H +\nu P}\rt)\Or\rt).
\eeq
It turns out that, in such states, simply as a consequence of the conservation laws \eqref{conslaw} and using clustering, certain symmetry relations hold for susceptibilities (these may be shown using arguments presented in \cite{D_2014}):
\beq
	\int \dd x\,\big(\bra h(x)k(0)\ket - \bra h(x)\ket\bra k(0)\ket\big)
	= \int \dd x\,\big(\bra p(x) j(0)\ket - \bra p(x)\ket\bra j(0)\ket\big).
\eeq
That is, the change of a current associated to a conservation law, under the variation of the potential associated to a different conservation law, is equal to the same quantity with the conservation laws exchanged:
\beq
	\frc{\p}{\p\beta} \bra k\ket = \frc{\p}{\p\nu}\bra j\ket.
\eeq
This is equivalent to the existence a (differentiable) potential ${\cal J}(\beta,\nu)$ that reproduces the averages currents:
\beq
	\bra j\ket = \frc{\p}{\p\beta} {\cal J}(\beta,\nu),\quad
	\bra k\ket = \frc{\p}{\p\nu} {\cal J}(\beta,\nu),
\eeq
generalizing the free energy that reproduces the average densities $\bra h\ket$ and $\bra p \ket$.

Both of the above families of relations can of course be generalized to the presence of more than two conserved currents.

\section{Open questions}

Let us conclude by enumerating a few open problems, some of them doable in a short time, others certainly more difficult.
\begin{itemize}

\item The construction of non-equilibrium steady states we just described is grounded on CFT. Since out-of-equilibrium phenomena may be sensitive to high-energy states, it is natural to wonder whether CFT is directly applicable to out-of-equilibrium physics in gapless systems. We have argued that the effects of large-energy states must diffuse away, but fully answering this question would for instance require a deeper analysis of the interplay between the low-energy and long-time limits. This is clearly a difficult question whose answer may depend on which out-of-equilibrium phenomena we are aiming at describing. However, the ballistic character of transport in samples of size smaller than both the mean free path $\ell_e$ and the phase decoherence length $L_\phi$ hints towards the validity of the CFT approach. The experimental observation \cite{Pierre_et_al} of the $T^2$ dependance of the mean energy current also provides arguments in favor of the validity of CFT approach, as long as one looks for transport properties in mesoscopic samples.

\item Transport ceases to be ballistic and becomes diffusive at large enough times and for samples of sizes comparable to or bigger than the mean free path $\ell_e$, even if coherent effects are not suppressed (i.e. in the case where the mean free path is smaller than the phase decoherence length). To our knowledge there is no simple model describing this crossover, in particular from ballistic non-equilibrium CFT to diffusive transport. We however recently made some progress in this direction \cite{BDavenir}. A related question is as to the crossover that is expected to occur in QFT models at weak couplings, between a free-particle pre-relaxation regime and an interaction-driven hydrodynamic steady state.

\item As we have described, non-equilibrium CFT can be extended to systems where defects partially transmit and reflect energy flows. Although this construction gives an arguably implicit characterization of the steady states, only the mean energy current has been computed. There is yet no formula for the large deviation functions for energy transfer through conformal defects (except for free-fermion models). Obtaining these formulas is an important step towards understanding general rules coding for non-equilibrium states in extended quantum systems. Even the simpler fact that the mean energy current is proportional to the difference of the square of the temperatures has not yet been algebraically proven within the framework of non-equilibrium CFT  in the presence of defects.

\item Similarly, the hydrodynamics approach should be extended to include the effects of defects, and to the study of the ballistic-to-diffusive crossover (see however \cite{BDavenir} for recent progresses).

\item The heuristic arguments of the hydrodynamic approach pointed towards the definition of random flows whose statistical properties are defined through the correlation functions of conformal fields (in the present case these fields were the stress tensor components). It will be interesting to try to develop a more precise, if not a more rigorous, definition of these random flows.

\item The construction of non-equilibrium steady states we have described bares similarities with the recently developed `macroscopic fluctuation theory' (MFT) \cite{MFT,Bodineau-Derrida} applicable to a large class of classical out-of-equilibrium systems. This is particularly apparent in the hydrodynamic approach based on boosted local density matrices. It seems important to decipher the analogies between these two constructions, in order to make one step towards developing a quantum version of the macroscopic fluctuation theory.

\item In higher-dimensional CFT, the new and fast developing techniques of holography, whereby the CFT is described as the asymptotic-boundary degrees of freedom of a gravity theory, are also particularly well adapted to non-equilibrium problems. In a particular limit (the ``large-$N$'' limit), classical gravity is involved, and can be seen as an extension of the hydrodynamic equations for interacting QFT. This point of view has been used \cite{Nat-Phys} in order to derive the boosted form of the non-equilibrium density matrix, independently from hydrodynamic arguments. It was also used in order to verify numerically the validity of the resulting non-equilibrium steady state \cite{adsnumer}. It would be interesting to develop further these techniques, especially in the study of fluctuations.

\item Finally, the problem of obtaining exact energy currents and fluctuations in the partitioning approach for integrable quantum chains is still open. 
\end{itemize}

\medskip

{\bf Acknowledgments:} 
This work was supported in part by the French `Agence National de la Recherche (ANR)' contract ANR-14-CE25-0003-01.
Both authors thank the Isaac Newton Institute for Mathematical Sciences for hospitality, under grant number EP/K032208/1, where the writing of this review started.

\end{document}